\documentclass[epj,final]{svjour}
\usepackage{epsfig, graphicx}

\usepackage{caption}
\usepackage{subfig}
\usepackage{multirow}
\usepackage{color}
\usepackage{epstopdf}
\newcommand{\be}{\begin{eqnarray}}
\newcommand{\ee}{\end{eqnarray}}

\newcommand{\er}{$\pm$}
\newcommand{\bc}{\begin{center}}
\newcommand{\ec}{\end{center}}

\begin{document}

%
%

\title{Analysis of pion production data measured by HADES\\in proton-proton collisions at 1.25 GeV}

\author{G.~Agakishiev$^{7}$, A.~Balanda$^{3}$, D.~Belver$^{18}$, A.~Belyaev$^{7}$,
J.C.~Berger-Chen$^{10,9}$, A.~Blanco$^{2}$, M.~B\"{o}hmer$^{10}$, J.~L.~Boyard$^{16}$, P.~Cabanelas$^{18}$,
S.~Chernenko$^{7}$
% ...
}

\institute{
\mbox{$^{1}$Istituto Nazionale di Fisica Nucleare - Laboratori Nazionali del Sud, 95125~Catania, Italy}\\
\mbox{$^{2}$LIP-Laborat\'{o}rio de Instrumenta\c{c}\~{a}o e F\'{\i}sica Experimental de Part\'{\i}culas , 3004-516~Coimbra, Portugal}\\
\mbox{$^{3}$Smoluchowski Institute of Physics, Jagiellonian University of Cracow, 30-059~Krak\'{o}w, Poland}\\
\mbox{$^{4}$GSI Helmholtzzentrum f\"{u}r Schwerionenforschung GmbH, 64291~Darmstadt, Germany}\\
\mbox{$^{5}$Technische Universit\"{a}t Darmstadt, 64289~Darmstadt, Germany}\\
\mbox{$^{6}$Institut f\"{u}r Strahlenphysik, Helmholtz-Zentrum Dresden-Rossendorf, 01314~Dresden, Germany}\\
\mbox{$^{7}$Joint Institute of Nuclear Research, 141980~Dubna, Russia}\\
\mbox{$^{8}$Institut f\"{u}r Kernphysik, Goethe-Universit\"{a}t, 60438 ~Frankfurt, Germany}\\
\mbox{$^{9}$Excellence Cluster 'Origin and Structure of the Universe' , 85748~Garching, Germany}\\
\mbox{$^{10}$Physik Department E12, Technische Universit\"{a}t M\"{u}nchen, 85748~Garching, Germany}\\
\mbox{$^{11}$II.Physikalisches Institut, Justus Liebig Universit\"{a}t Giessen, 35392~Giessen, Germany}\\
\mbox{$^{12}$Istituto Nazionale di Fisica Nucleare, Sezione di Milano, 20133~Milano, Italy}\\
\mbox{$^{13}$Institute for Nuclear Research, Russian Academy of Science, 117312~Moscow, Russia}\\
\mbox{$^{14}$Institute of Theoretical and Experimental Physics, 117218~Moscow, Russia}\\
\mbox{$^{15}$Department of Physics, University of Cyprus, 1678~Nicosia, Cyprus}\\
\mbox{$^{16}$Institut de Physique Nucl\'{e}aire, CNRS-IN2P3, Univ. Paris-Sud, Universit\'{e} Paris-Saclay, 91406~Orsay Cedex, France}\\
\mbox{$^{17}$Nuclear Physics Institute, Academy of Sciences of Czech Republic, 25068~Rez, Czech Republic}\\
\mbox{$^{18}$LabCAF. F. F\'{\i}sica, Univ. de Santiago de Compostela, 15706~Santiago de Compostela, Spain}\\
\mbox{$^{19}$NRC "Kurchatov Institute", PNPI, 188300, Gatchina, Russia}\\
\\
\mbox{$^{a}$ also at Lawrence Berkeley National Laboratory, ~Berkeley, USA}\\
\mbox{$^{b}$ also at ISEC Coimbra, ~Coimbra, Portugal}\\
\mbox{$^{c}$ also at ExtreMe Matter Institute EMMI, 64291~Darmstadt, Germany}\\
\mbox{$^{d}$ also at Technische Universit\"{a}t Dresden, 01062~Dresden, Germany}\\
\mbox{$^{e}$ also at Dipartimento di Fisica, Universit\`{a} di Milano, 20133~Milano, Italy}\\
\mbox{$^{f}$ also at Frederick University, 1036~Nicosia, Cyprus}\\
\mbox{$^{g}$ also at Dipartimento di Fisica and INFN, Universit\`{a} di Torino, 10125~Torino, Italy}\\
\mbox{$^{h}$ also at Utrecht University, 3584 CC~Utrecht, The Netherlands}\\
\mbox{$^{i}$ also at Helmholtz--Institut f\"ur Strahlen-- und Kernphysik, Universit\"at Bonn, Germany}\\
}

\date{Received: date / Revised version: date}

\abstract{Baryon resonance production in proton-proton collisions at a kinetic beam energy of 1.25 GeV is investigated. The multi-differential data were measured by the HADES collaboration. Exclusive channels with one pion in the final state ($np\pi^{+}$ and $pp\pi^{0}$) were put to extended studies based on various observables in the framework of a one-pion exchange model and  with solutions obtained within the framework of a partial wave analysis (PWA) of the Bonn-Gatchina group. The results of the PWA confirm the dominant contribution of the $\Delta$(1232), yet with a sizable impact of the $N$(1440) and non-resonant partial waves.
\PACS{ {13.30.-a}{} \and 
       {13.60.Le}{} \and 
       {14.20.Gk}{} \and
       {11.80.Et}{}
     } 
}

\authorrunning{G.~Agakishiev et al.}
\titlerunning{p(1.25 GeV) + p $\rightarrow$ pp$\pi^{0}$, pn$\pi^{+}$}
\maketitle

\begingroup
\renewcommand{\thefootnote}{\alph{footnote}}
\footnotetext{$^{*}$ Corresponding authors: witold.przygoda@uj.edu.pl, andsar@hiskp.uni-bonn.de}
\endgroup


\section{Introduction}

Nucleon-nucleon interactions provide a fundamental knowledge about the nature of nuclear forces with a strong impact on the construction of many dynamic models. Although the elastic $NN$ scattering is a dominant process at the low energies, the understanding of inelastic collisions is mandatory above the one-pion and two-pion production thresholds (for a review see \cite{Moskal2002}). One of the first semi-phenomenological models by Mandelstam \cite{Mandelstam1958} was describing the pion production by the formation of the intermediate $N\Delta$ state and a decay of the $\Delta$ into a nucleon and a pion. However, the absence of dependence of the production amplitude on energy was in contradiction to experimental data at energies above 0.7 GeV. A more advanced approach was realized by one-pion (OPE, see \cite{Ferrary1963}) or by a one-boson exchange (OBE) models, developed by several groups (see i.e. \cite{Gersten1971,Bryan1972,Schierholz1972,Erkelenz1974,Holinde1975,Machleidt1987,Engel1996}). The improved version of the OPE model was proposed by Suslenko and Gaisak \cite{Suslenko1986}, describing the experimental data in the $pp \rightarrow np\pi^{+}$ reaction in the energy range $0.6-1.0$ GeV with an accuracy of $10-15\%$. The model was tested also in the $pp \rightarrow pp\pi^{0}$ reaction at seven energies of the incident proton in the range $0.6-0.9$ GeV with the data collected at the PNPI \cite{Andreev1994}. Although various differential distributions are described by the model qualitatively well, the predicted total cross sections are lower than the reconstructed from the experimental data. Similar conclusions were reported in the study of the neutral pion production at proton beam momentum 1.581 GeV/c (kinetic energy 0,9 GeV) and 1.628 GeV/c (kinetic energy 0.9089 GeV) \cite{Sarantsev2004}, supplemented by the recent studies of $np\pi^{+}$ channel (\cite{Ermakov2014} and \cite{Ermakov2011}) for the same beam momenta. The good data description in the $np\pi^{+}$ channel leads to the underestimation of the total cross section by the OPE model \cite{Suslenko1986} in the $pp\pi^{0}$ channel.

Yet another OPE model, successfully describing the data at slightly higher energies of 0.97 GeV \cite{Bugg1964}, 1.48 GeV \cite{Eisner1965} and 4 GeV \cite{Coletti1967} in the $pp \rightarrow np\pi^{+}$ channel,  was introduced by Dmitriev \textit{et al.} \cite{Dmitriev1986}. The matrix element in the model is calculated based on the direct and exchange graphs for $\Delta$-production in $pp$ collisions, with the form factor in the $\pi NN$ and $\pi N\Delta$ vertices
\begin{equation}
F(t)=\frac{\Lambda_{\pi}^{2} - m_{\pi}^{2}}{\Lambda_{\pi}^{2} - t}
\label{e_DMITRIEV}
\end{equation}
where $\Lambda_{\pi}$ is the coupling constant adjusted to the data (i.e. $\Lambda_{\pi} = 0.63$ GeV for the \cite{Bugg1964}) and $t$ denotes the Mandelstam variable for the momentum transfer. 

A more versatile dynamical model by Teis \textit{et al.} \cite{Teis1997} describes the production of light mesons in proton-proton collisions and extends it to heavy-ion collisions in the energy range of $1-2$ GeV/u. The major assumptions of this model are: (i) the entire meson production proceeds via intermediate resonance excitation (ii) the total cross section amounts to the incoherent sum of all resonances contributing to a specific channel.
The matrix elements for the resonance production were obtained from a fit to the data of 1$\pi$, $\eta$, $\rho$ and 2$\pi$ production cross sections in nucleon-nucleon reactions. They were assumed to be constant, except for the $\Delta$ where dependency on $t$ was adopted from \cite{Dmitriev1986}. A similar approach is also used in other resonance models, e.g. GiBUU \cite{Buss2012,Weil2012} (with only small modifications of the Teis model \cite{Teis1997}) and UrQMD \cite{Bass1998}.

The modelling of the angular distributions of the produced resonances allows for a more detailed comparison with experimental data and is essential when measurements within a limited acceptance coverage are considered. For example, the OPE model of Dmitriev \textit{et al.} \cite{Dmitriev1986} provides anisotropic angular description of the $\Delta$ resonance in accordance with experimental data. Other important observables characterizing a source of pion production are the various angular distributions in the nucleon-pion reference systems, i.e. helicity and Gottfried-Jackson frames \cite{Gottfried1964}. For instance, the angular distribution of the $\Delta$ decay depends on the population of different spin states excited in the $NN \rightarrow N\Delta$ process, what can be described in terms of a $4 \times 4$ spin density matrix $\rho_{ij}$. Integrating over the full azimuthal range and assuming solely a one-pion exchange, the decay angular distribution $\Delta\rightarrow N\pi$ follows a $\sim (1+3cos^2\theta)$ distribution, where $\theta$ is the angle of a pion (or a nucleon) in the $\Delta$ rest frame with respect to the beam axis (see \cite{Gottfried1964}). Such a parameterization is corroborated by the experimental data \cite{Andreev1994}. However, there is not much information on higher mass resonances, and usually isotropic distributions are used in resonance models.

In view of the limitations of the resonance model \cite{Teis1997}, the partial wave analysis provides a significant advantage - it includes the coherent sum of both resonant and non-resonant contributions within the solution based on the simultaneous fit to many experimental data samples. The extraction of contributions from different partial waves is performed event by event and based on the maxium-likelihood method. The angular distributions for a given partial wave in the final state, characterized by the spin and parity, are naturally accounted for. Therefore, resonant and non-resonant contributions, including interferences, can be extracted. In this work we compare HADES results on one-pion production obtained with calculations based on the resonance model of \cite{Teis1997} and of the partial wave analysis developed by the Bonn-Gatchina group \cite{Ermakov2011}.

A detailed knowledge of the resonance production is also essential for the understanding of dielectron sources in nucleon and pion induced reaction (see \cite{Hades2014}). In particular the $\Delta\rightarrow Ne^+e^-$  Dalitz decay presents the next, after the neutral pion decay, important source of lepton pairs at beam energies around 1 GeV. The corresponding branching ratio of the decay and its dependence on the dielectron invariant mass have not yet been measured in the exclusive process like, for example, $pp\rightarrow ppe^+e^-$. The analysis of the hadronic channel $pp\rightarrow\Delta^+p\rightarrow pp\pi^0$, presented in this work, provides a $\Delta+$ resonance contribution, being a mandatory reference to deduce the branching ratio for the dilepton decay of $\Delta \to Ne^+e^-$ (a subject of a forthcoming publication).

Our paper is organized as follows: Section \ref{thehadesexp} introduces the experimental set-up, conditions under which $np\pi^+$ and $pp\pi^0$ channels were selected and the normalization procedure. The channels are analyzed within the resonance model ansatz assuming the excitation of $\Delta$(1232) and $N$(1440). Various differential distributions within the HADES acceptance are presented and compared to calculations in Section \ref{OPE_results}. The acceptance corrected differential and total cross sections are shown in subsections \ref{nppip_channel} and \ref{pppi0_channel}. Section \ref{PWA_results} presents the methodology of the partial wave analysis in $NN$ collisions and the discussion of the obtained solutions: contributing partial waves, the role of the resonances as well as the non-resonant contributions. Finally, the experimental data are acceptance corrected with the PWA solution. The conclusions compare results obtained with the two methods.

\section{The HADES experiment}
\label{thehadesexp}

The experiment was performed with the High Acceptance Di-Electron Spectrometer (HADES) \cite{Hades2009}  installed at the GSI Helm\-holtz\-zentrum f\"{u}r Schwerionenforschung in Germany. A proton beam of $10^7$ particles/s with a kinetic energy of $1.25$ GeV was incident on a liquid hydrogen target. The analysis of this experiment was described already in \cite{Ramstein2012}. In this report we extend the studies on various additional observables: angular projections of the identified particles, invariant masses as well as the angular projections in the helicity and in the Gottfried-Jackson frames. 

To study one-pion production mechanisms in the hadronic channels, only events with one proton and one pion ($p\pi^{+}$) and two protons ($pp$) were identified with the help of the missing mass technique. The collected statistic amounts to $2.73 \times 10^6$ events with an identified $\pi^{+}$ and $0.53 \times 10^6$ events with an identified $\pi^{0}$, respectively. The background estimation was done on the base of a double-differential missing mass spectrum obtained for 20 different bins in the variable $cos~\theta^{CM}_{\pi N}$ and 25 bins in $M^{inv}_{\pi N}$. Prior to the background evaluation, the two-pion contribution to the missing mass spectrum, not very sensitive to details of the two-pion production model, was subtracted, as explained in \cite{Ramstein2012}. The background contribution obtained from the fit procedure applied to the $np\pi^{+}$ final state amounts to a few percents. In the case of the $pp\pi^{0}$ sample the background contribution yields to about ten percents. The estimated background is used to calculate the ratio of signal to total yields, utilized as weights (Q-factors) in the event-by-event partial wave analysis. 

\begin{table*}
\centering
\begin{tabular}{|l|l|c|c|}
\hline 
\multicolumn{1}{|c|}{\bf{final state}} &
\multicolumn{1}{|c|}{\bf{intermediate process}} & \bf{$\sigma_{RES}$ (mb) } & \bf{$\sigma_{PWA}$ (mb) }     \\
\hline $np\pi^{+}$ & $pp\rightarrow n\Delta^{++}(1232)$ & 16.90 & 11.1~\er~0.4       \\
\cline{2-4} & $pp\rightarrow p\Delta^{+}(1232)$  & ~1.89 & 1.2~\er~0.2        \\
\cline{2-4} & $pp\rightarrow pN(1440)$ & ~0.54 & ~1.7~\er~0.2            \\
\cline{2-4} & \multicolumn{1}{|r|}{$Total:$} & 19.35 & ~16.34~\er~0.8            \\
\hline $pp\pi^{0}$ & $pp\rightarrow p\Delta^{+}(1232)$  & ~3.76 &~2.96~\er~0.07     \\
\cline{2-4} & $pp\rightarrow pN(1440)$ & ~0.27 & ~0.86~\er~0.06            \\
\cline{2-4} & \multicolumn{1}{|r|}{$Total:$} & ~4.03 & ~4.2~\er~0.15          \\
\hline
\end{tabular}
\caption {Cross sections for the $p (1.25~GeV) + p$ reaction and one-pion final states with the intermediate baryon resonance excitation: $\sigma_{RES}$ for the resonance model \cite{Teis1997}, $\sigma_{PWA}$ for the partial wave fit.}
\label{cross_sections_tab}
\end{table*}

\begin{figure*}
            \centering
                    \includegraphics[width=0.3\textwidth]{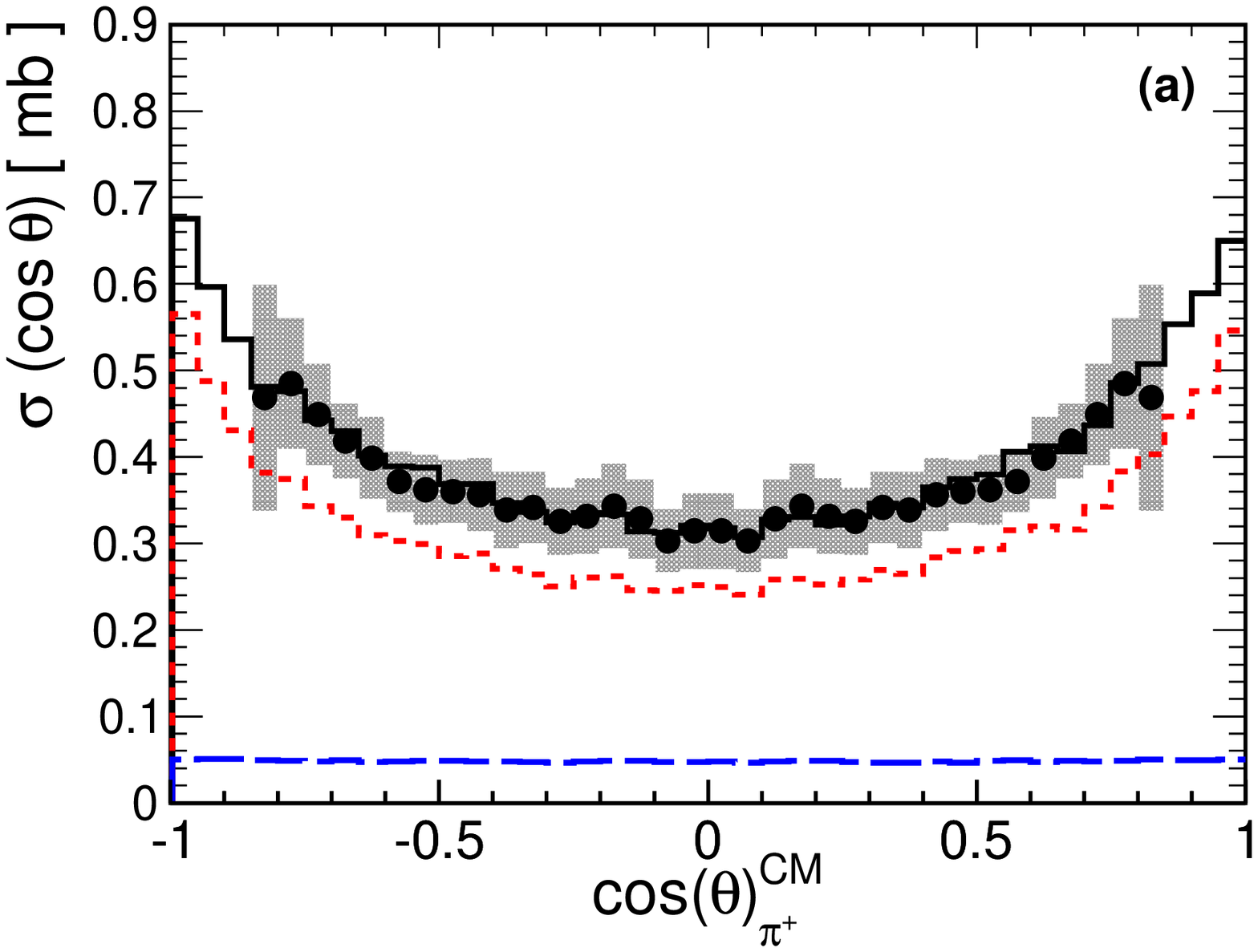}
                    \includegraphics[width=0.3\textwidth]{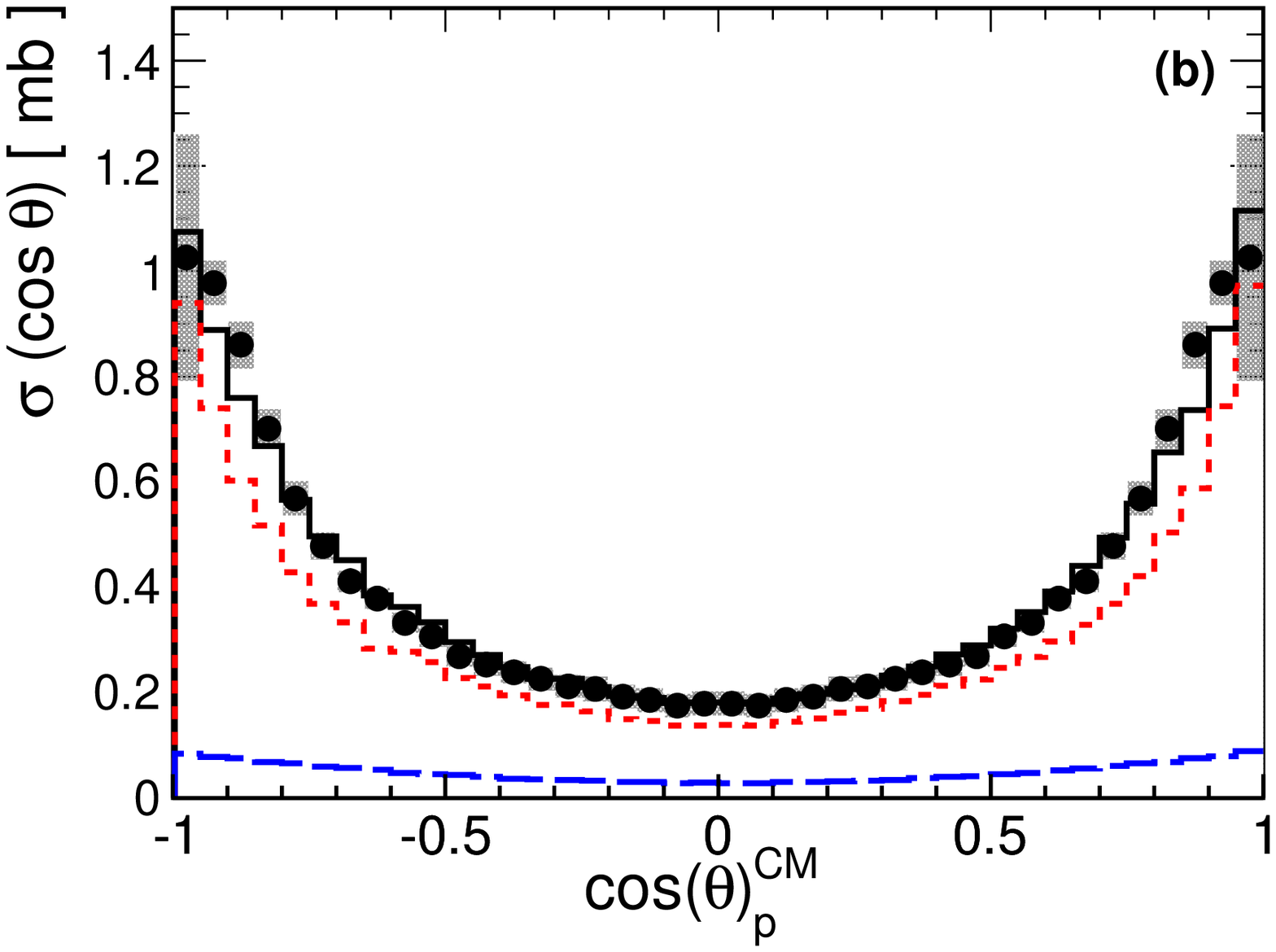}
                    \includegraphics[width=0.3\textwidth]{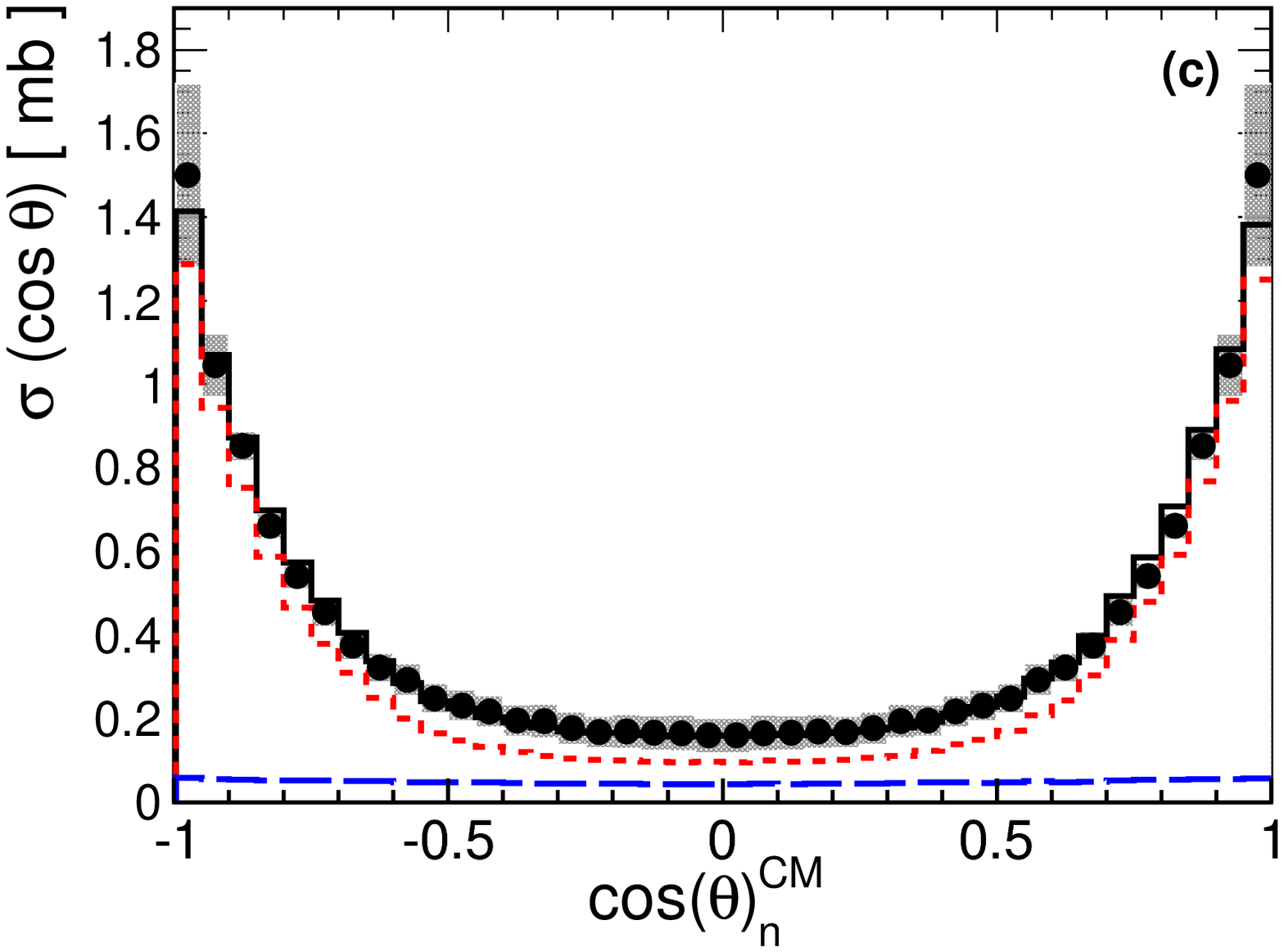}\\
                    Angular distribution of (a) $\pi^{+}$, (b) $p$ and (c) $n$ in c.m.s. reference frame.\\

                    \includegraphics[width=0.3\textwidth]{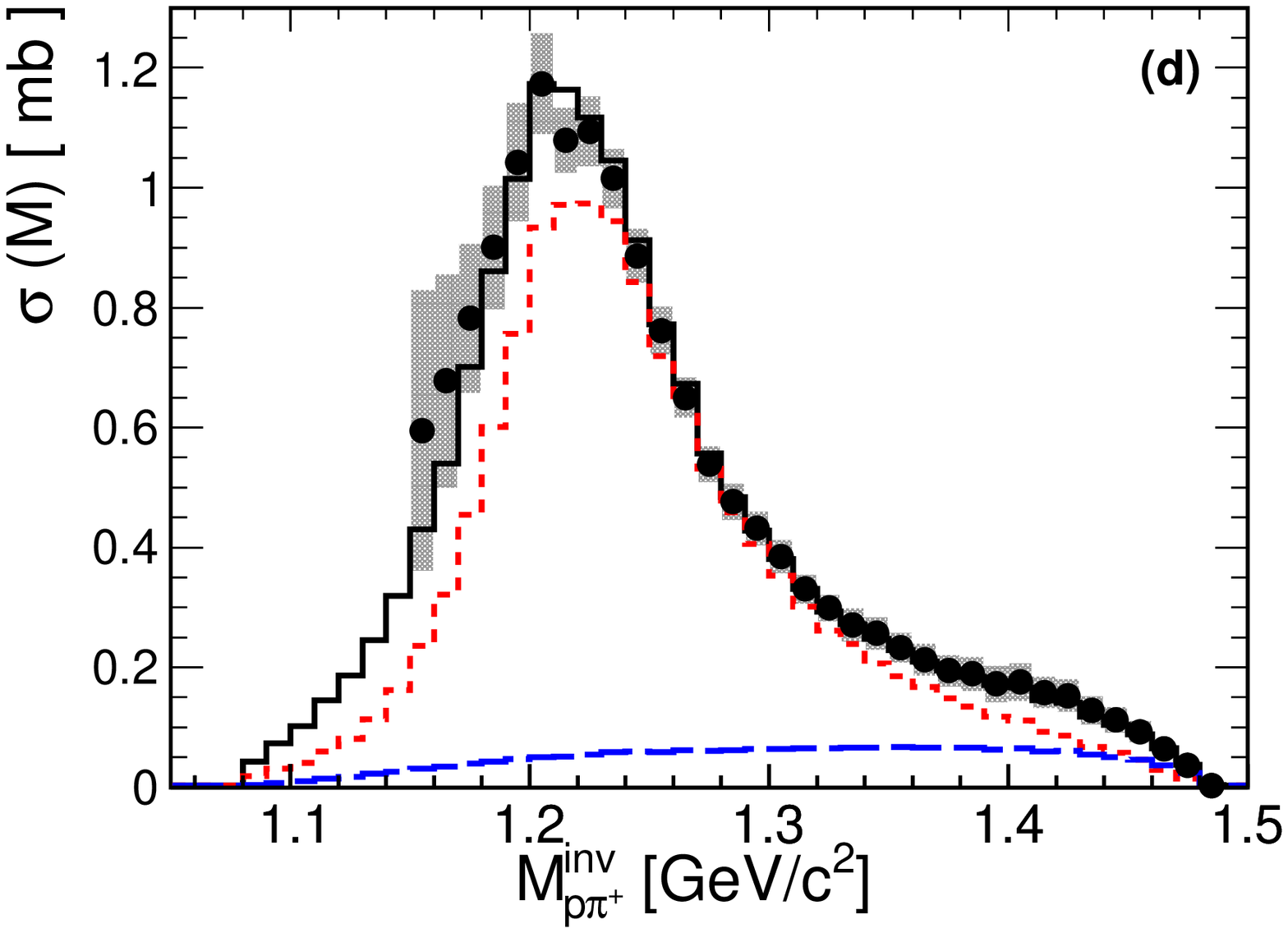}
                    \includegraphics[width=0.3\textwidth]{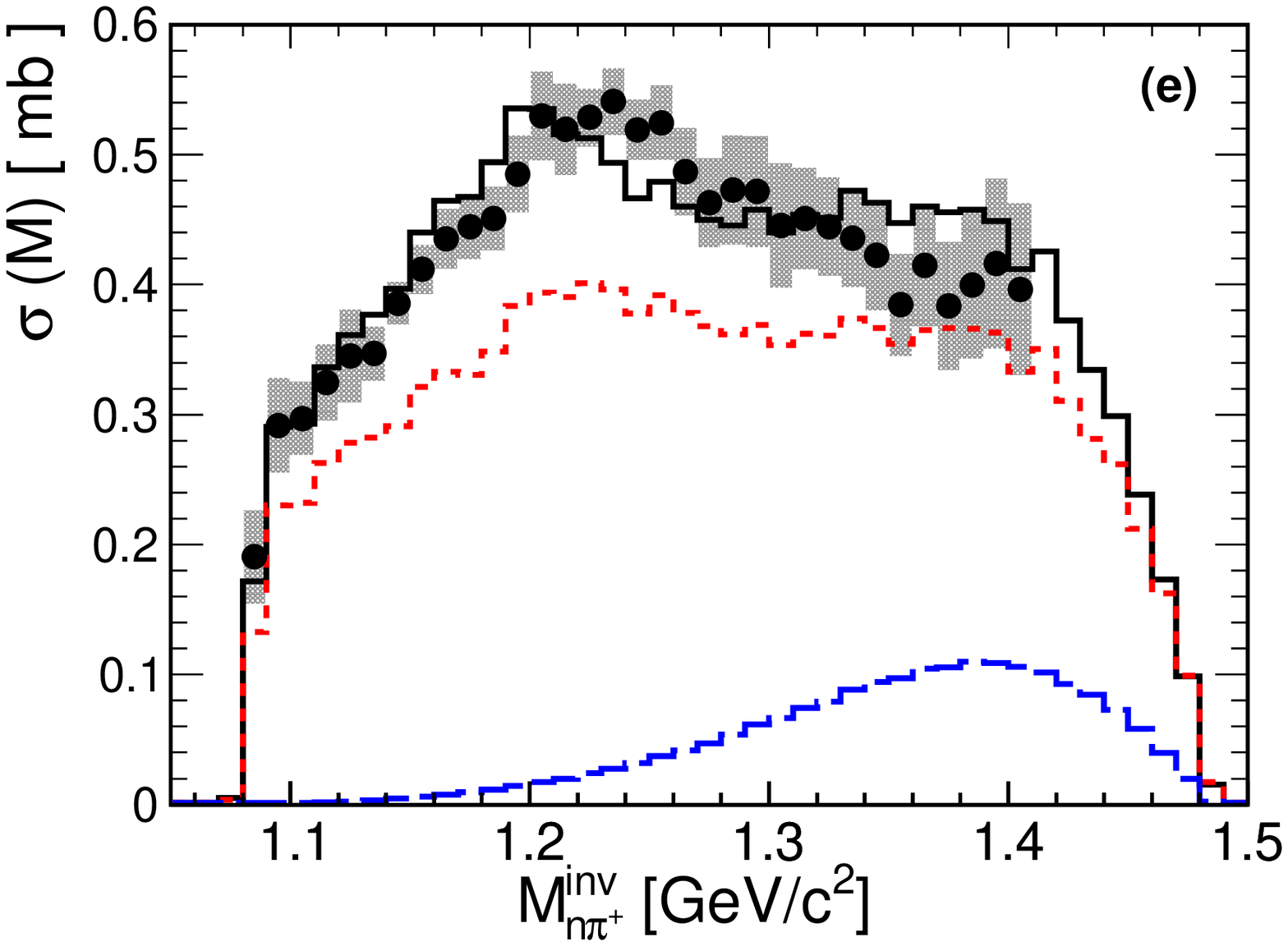}
                    \includegraphics[width=0.3\textwidth]{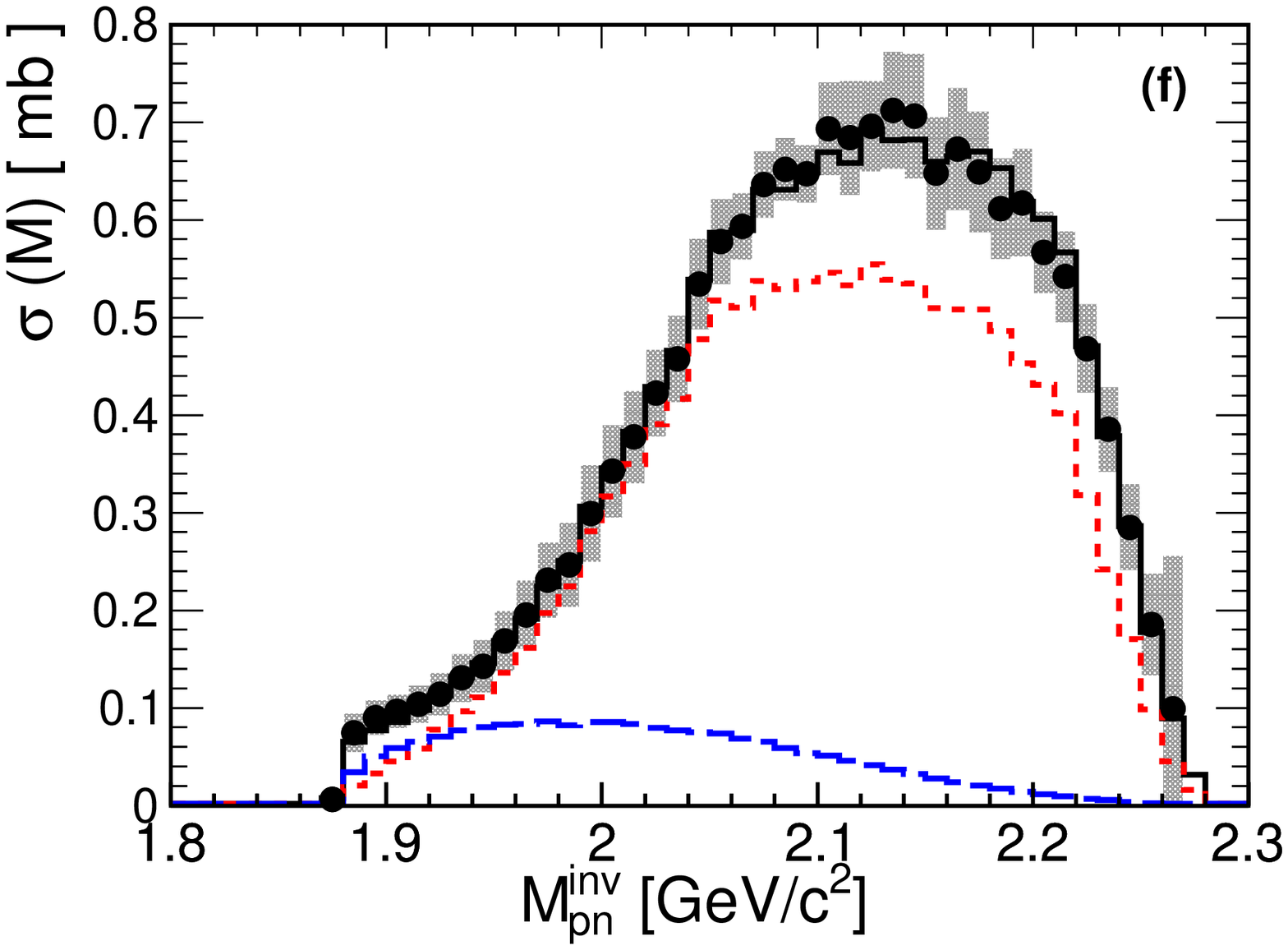}\\
                    Invariant mass of (d) $p\pi^{+}$, (e) $n\pi^{+}$ and (f) $pn$.\\

                    \includegraphics[width=0.3\textwidth]{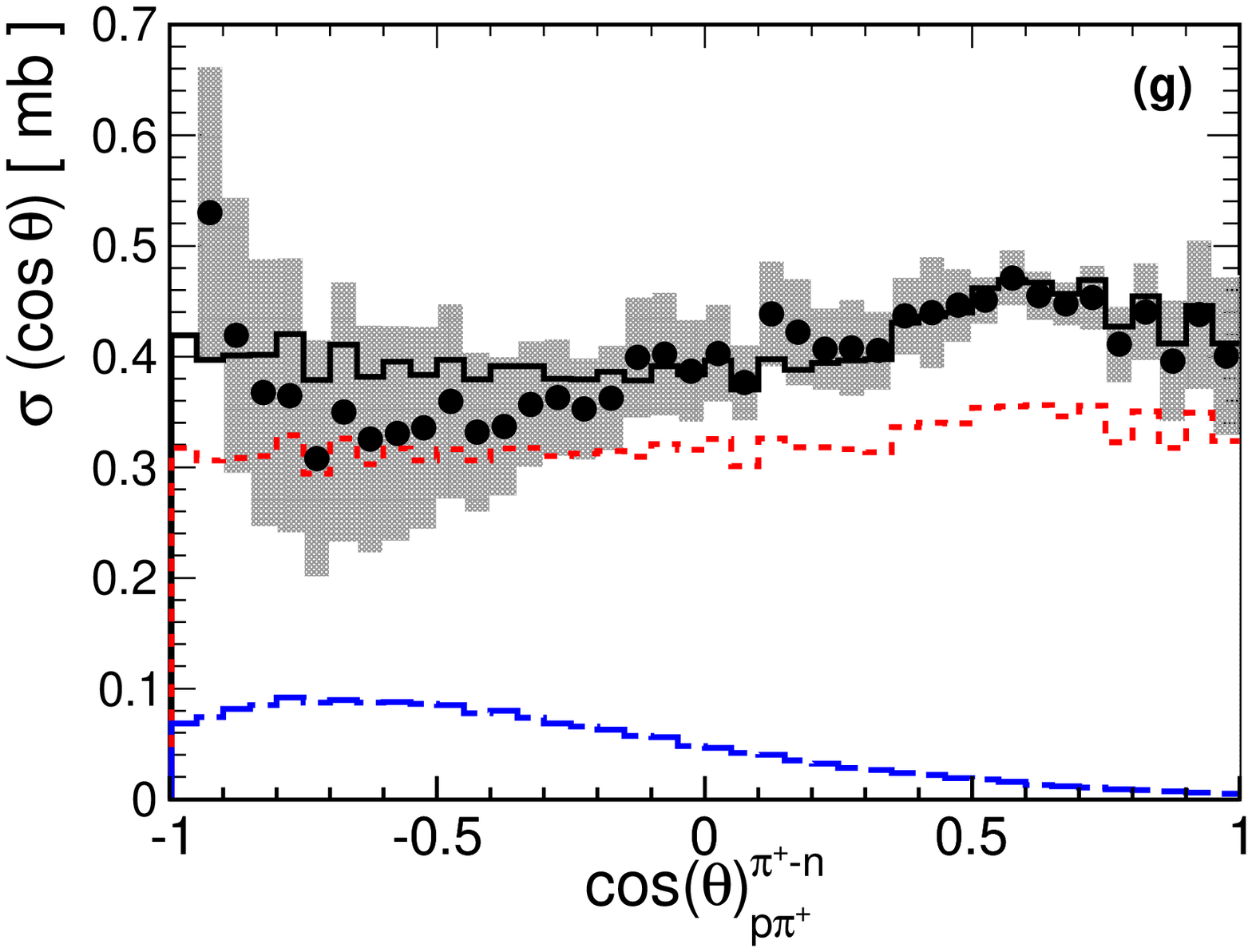}
                    \includegraphics[width=0.3\textwidth]{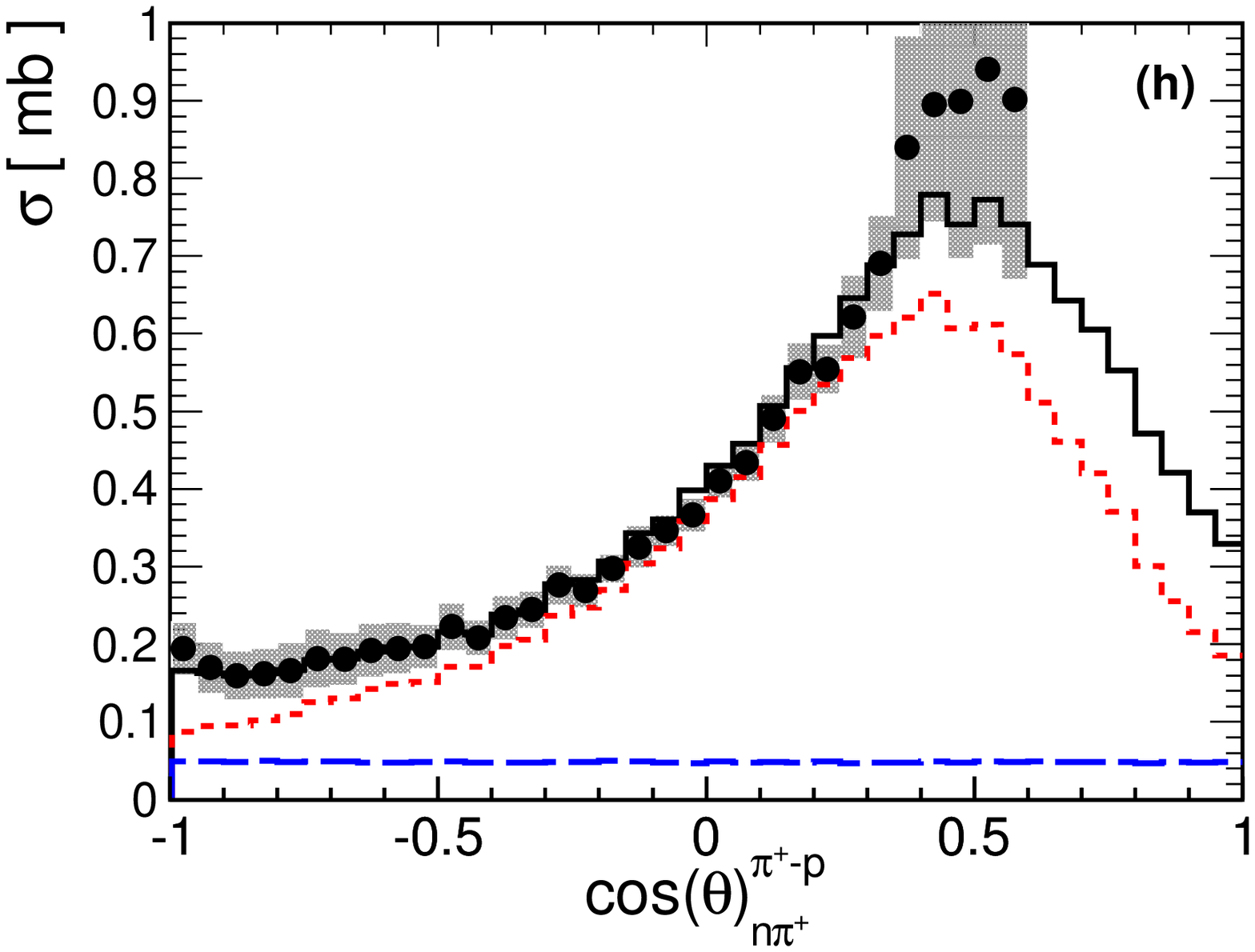}
                    \includegraphics[width=0.3\textwidth]{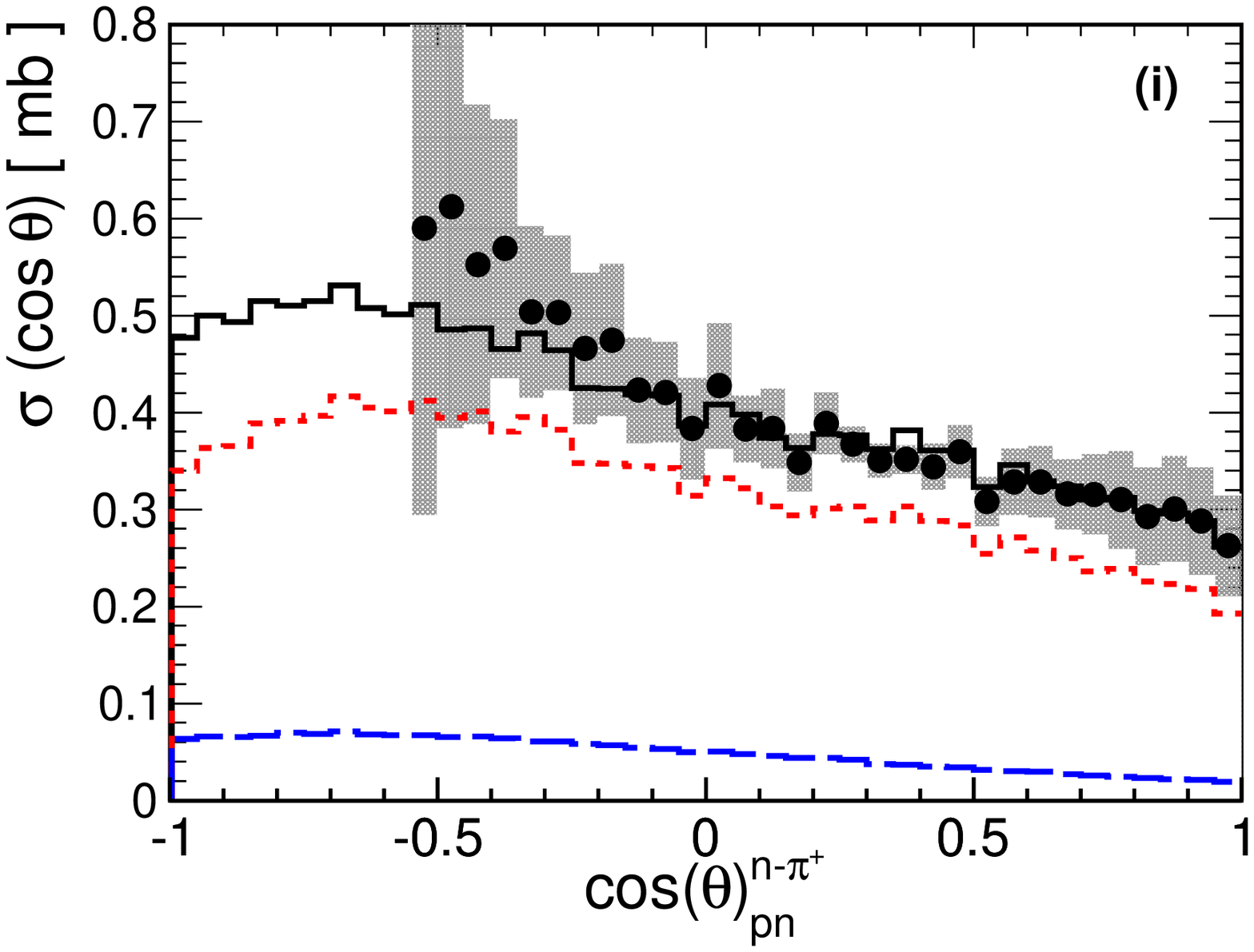}\\
                    Helicity distribution of (g) $\pi^{+}$ in $p\pi^{+}$ reference frame, (h) $\pi^{+}$ in $n\pi^{+}$ reference frame and (i) $n$ in $pn$ reference frame.\\
            
                    \includegraphics[width=0.3\textwidth]{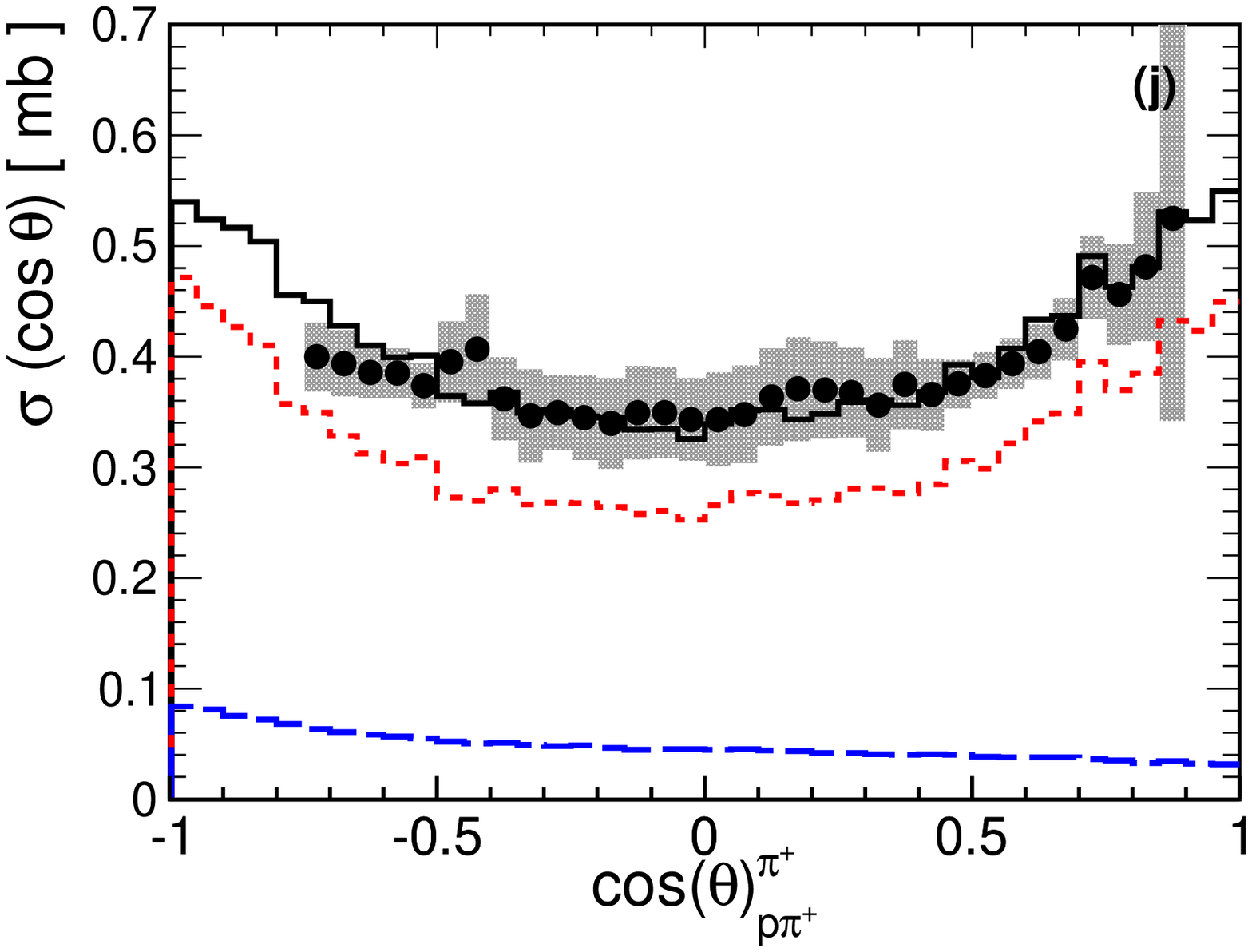}
                    \includegraphics[width=0.3\textwidth]{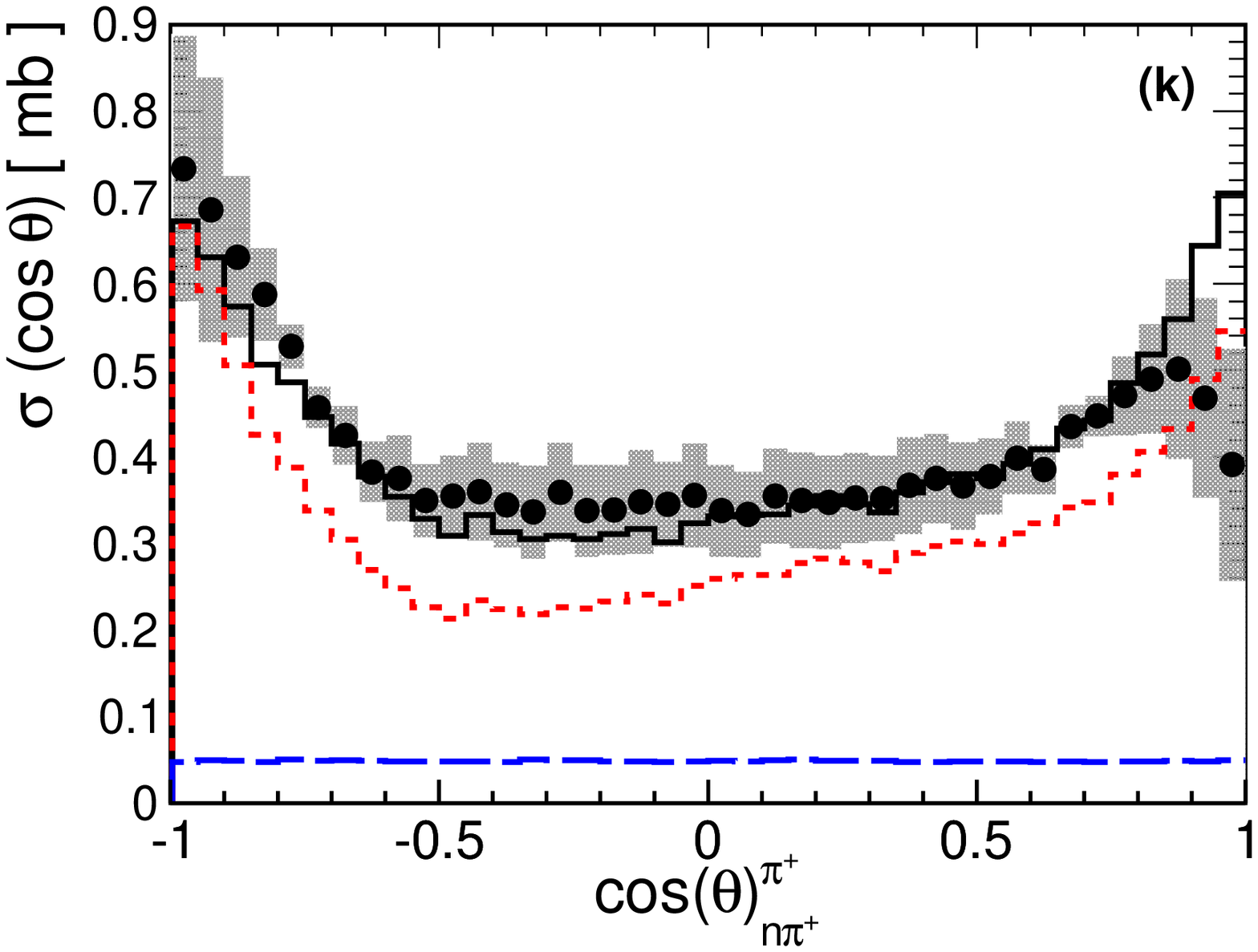}
                    \includegraphics[width=0.3\textwidth]{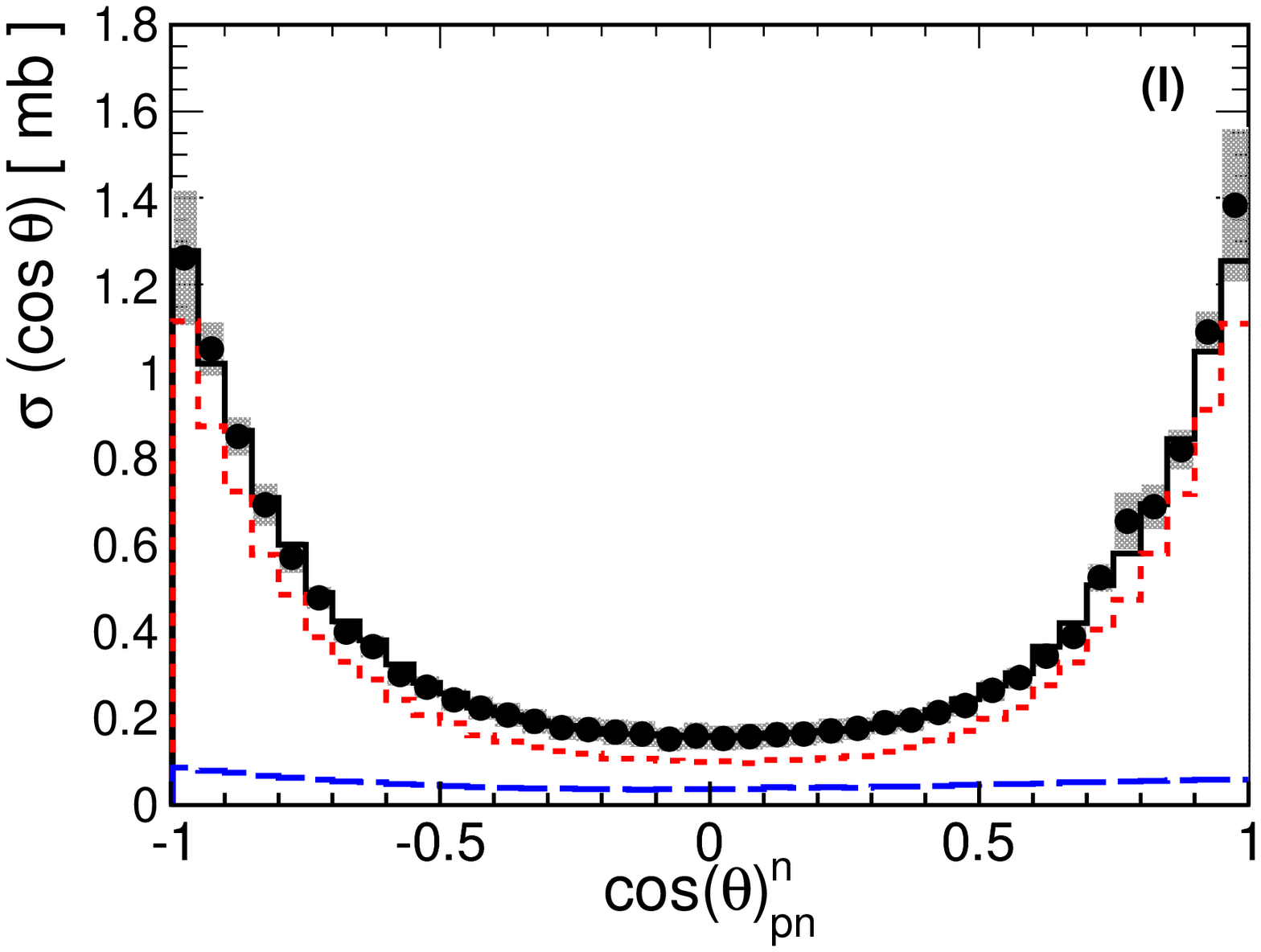}\\
                    Angular distribution of (j) $\pi^{+}$ in $p\pi^{+}$ GJ reference frame, (k) $\pi^{+}$ in $n\pi^{+}$ GJ reference frame and (l) $n$ in $pn$ GJ reference frame.\\

            \caption{(Color online) Various projections for the $np\pi^{+}$ channel: uncorrected data points (black) within the HADES acceptance with systematic and statistical vertical error bars, normalized to the number of $pp$ elastic scattering ($N_{el}$). Histograms: total PWA solution folded within the HADES acceptance and efficiency (solid black) and normalized to the respective yields of experimental data, the $\Delta$(1232) contribution (short-dashed red) and the $N$(1440) contribution (long-dashed blue). Dotted histogram (black): modified resonance model. The grey hatched area in each panel shows the distribution in the case of isotropically simulated particles.}

    \label{f1}
    \end{figure*}

All spectra presented in Figs. \ref{f1} and \ref{f2} are uncorrected distributions within the HADES acceptance. They are normalized to the $pp$ elastic scattering yield measured in the same experimental run. The reference $pp$ elastic cross section for the proton in the polar angle range between $46^{\circ}-135^{\circ}$ in c.m.s. amounts to 3.99~\er~0.23 mb (EDDA Collaboration \cite{EDDA2004}). The normalization error is estimated to be $8\%$, where $5.8\%$ is derived from the error of the reference differential cross section and $6\%$ is the systematic error of the reconstruction of events with elastic scattering (see \cite{Ramstein2012} for details).

\section{Results and comparison to resonance model}
\label{OPE_results}
To describe the data from the $p+p$ reaction, the resonance model by Teis \textit{et~al.} \cite{Teis1997} was implemented in the framework of the PLUTO event generator \cite{PLUTO2007}. Then, the full GEANT simulation and the Monte-Carlo simulations of the detector response, followed by the same analysis steps employed for the experimental data, were performed. The following hadronic channels were included: $pp \rightarrow$ (i) $n+\Delta^{++}(1232)$ with decay (BR = 1) $\Delta^{++} \rightarrow p+\pi^{+}$, (ii) $p+\Delta^{+}(1232)$ with subsequent decays (BR = 1/3) $\Delta^{+} \rightarrow n+\pi^{+}$  and (BR = 2/3) $\Delta^{+} \rightarrow p+\pi^{0}$, (iii) $p+N(1440)$ with decays (BR = 0.65*2/3) $N(1440) \rightarrow n+\pi^{+}$ and (BR = 0.65*1/3) $N(1440) \rightarrow p+\pi^{0}$ for the $NN$ $\rightarrow$ $N\Delta$(1232) reaction. The simulation employs the model of Dmitriev \textit{et~al.} \cite{Dmitriev1986} but replaces, as in the resonance model \cite{Teis1997}, the original parameterization of the $\Delta$ resonance total width by the one given in the Moniz model \cite{Koch1984}:

\begin{equation}
\Gamma(m)=\Gamma_{R}\frac{m_{R}}{m}\big(\frac{q}{q_{R}}\big)^{2l+1}\big(\frac{q_{R}^2+\delta^2}{q^2+\delta^2}\big)^{l+1}.
\label{e_MONIZ}
\end{equation}
$m_{R}$ and $\Gamma_{R}$ are the pole mass and the width of a resonance $R$, $m$ is the current resonance mass, $q$ and $q_{R}$ are the three-momenta of the pion in the reference frame of a resonance with mass $m$ and $m_{R}$, $l$ is the angular momentum of the emitted pion ($l=1$ for the $\Delta$). The quantity $\delta$ is a parameter in the cut-off function which, in the case of $\Delta$ resonance, equals to $\delta$ = 0.3 GeV/c \cite{Moniz1981}. Such a parameterization which suppresses the high-mass tail of the resonance, is compatible with the description of the HADES data at a higher energy \cite{Hades2014}. The parameterization of the one-pion decay width for the Roper resonance is defined in the similar manner (see \cite{Teis1997} for details). The final state interaction (FSI) between the outcoming nucleons was also modeled according to the Jost parameterization \cite{Titov2000}. 

The production cross sections for the intermediate resonances were also taken from the model \cite{Teis1997}, except for the Roper resonance, where a slightly larger cross section was used, based on a lagrangian model \cite{Cao2010}. Decay branching ratios to one and two pions at resonance pole masses are taken from the PDG review \cite{PDG2014}). Isospin relations lead to the following ratios between cross sections:
\begin{equation}
\sigma_{pp \rightarrow np\pi^+} = 5 \sigma_{pp \rightarrow pp\pi^0}
\label{e_CSRATIODELTA}
\end{equation}
for the $\Delta$ resonance with the isospin $I=\frac{3}{2}$, and 
\begin{equation}
\sigma_{pp \rightarrow pp\pi^0} = 2 \sigma_{pp \rightarrow np\pi^+}
\label{e_CSRATIOROPER}
\end{equation}
for the $N$(1440) (Roper) resonance with the isospin $I=\frac{1}{2}$. The cross sections are listed in Table~\ref{cross_sections_tab} (column $\sigma_{RES}$); the subsequent contributions to the total cross section were added incoherently. It is worth mentioning that the changes affect only the angular distributions of the $\Delta \to \pi N$ decay at large c.m.s. angles, keeping the cross section untouched. The calculations with the OPE model \cite{Dmitriev1986} remain still valid for most applications and are utilized successfully in modern resonance models (e.g. GiBUU \cite{Buss2012}).

\subsection{\bf{$np\pi^{+}$} channel}
\label{nppip_channel}

The description of the intermediate $\Delta^{++}$ resonance in the $pp \rightarrow np\pi^{+}$ channel within the OPE model \cite{Dmitriev1986} required the adjustment of the cut-off parameter $\Lambda_{\pi}$ in Eq. ($\ref{e_DMITRIEV}$) for the vertex form factor. The HADES data favour $\Lambda_{\pi}=0.75$ GeV (see \cite{Ramstein2012}). Further improvement could be achieved with the empirical parameterization of the angular distribution $cos \theta^{CM}_{p\pi^{+}}$ as a function of $M^{inv}_{p\pi^{+}}$. It allows to describe the anisotropic production of the resonance as a function of the invariant mass, in agreement with the observations of the former proton-proton experiments \cite{Colton1971}. The comparison of the improved model with data is shown in Fig.~\ref{f1}, where various projections of the uncorrected data and the Monte-Carlo simulation (dotted histogram), within the HADES acceptance, are presented. We show single particle angular distributions in the center of mass (c.m.s.), helicity and Gottfried-Jackson (GJ) frames and two-particle  invariant mass spectra. The calculations with the resonance model are compared to the data using the normalization explained above, while the results obtained with a partial wave analysis (explained in Section \ref{PWA_results}) and phase space distributions are normalized to the yield of the data.

Thanks to a good solid angle coverage and a good model description, the data could be corrected for the reconstruction inefficiencies and the detector acceptance, each distribution with the respective one-dimensional correction function. The correction function is constructed, for a given distribution, as ratio of the model yield in $4\pi$ and the yield within the HADES acceptance, including all detection and reconstruction inefficiencies obtained using the full analysis chain. The integrated correction factor in the $np\pi^{+}$ channel varies in the range 20-40, depending on the distribution. The extracted cross section for the $np\pi^{+}$ channel, in agreement with \cite{Ramstein2012}, amounts to $17.0\pm2.2$ (systematic error) mb, with a negligible statistical error. The systematic error includes: $5\%$ error due to the particle identification (selection cuts and the missing mass cut) and the background subtraction, $10\%$ error due to the correction and model uncertainty and $8\%$ is the normalization error (errors are added quadratically). The background subtraction error was deduced by varying of a polynomial function used together with a Gauss function to fit the missing mass spectrum. The model error was estimated from the differences in the integrated yields of the various distributions obtained after acceptance corrections.

\begin{figure*}
            \centering
                    \includegraphics[width=0.24\textwidth]{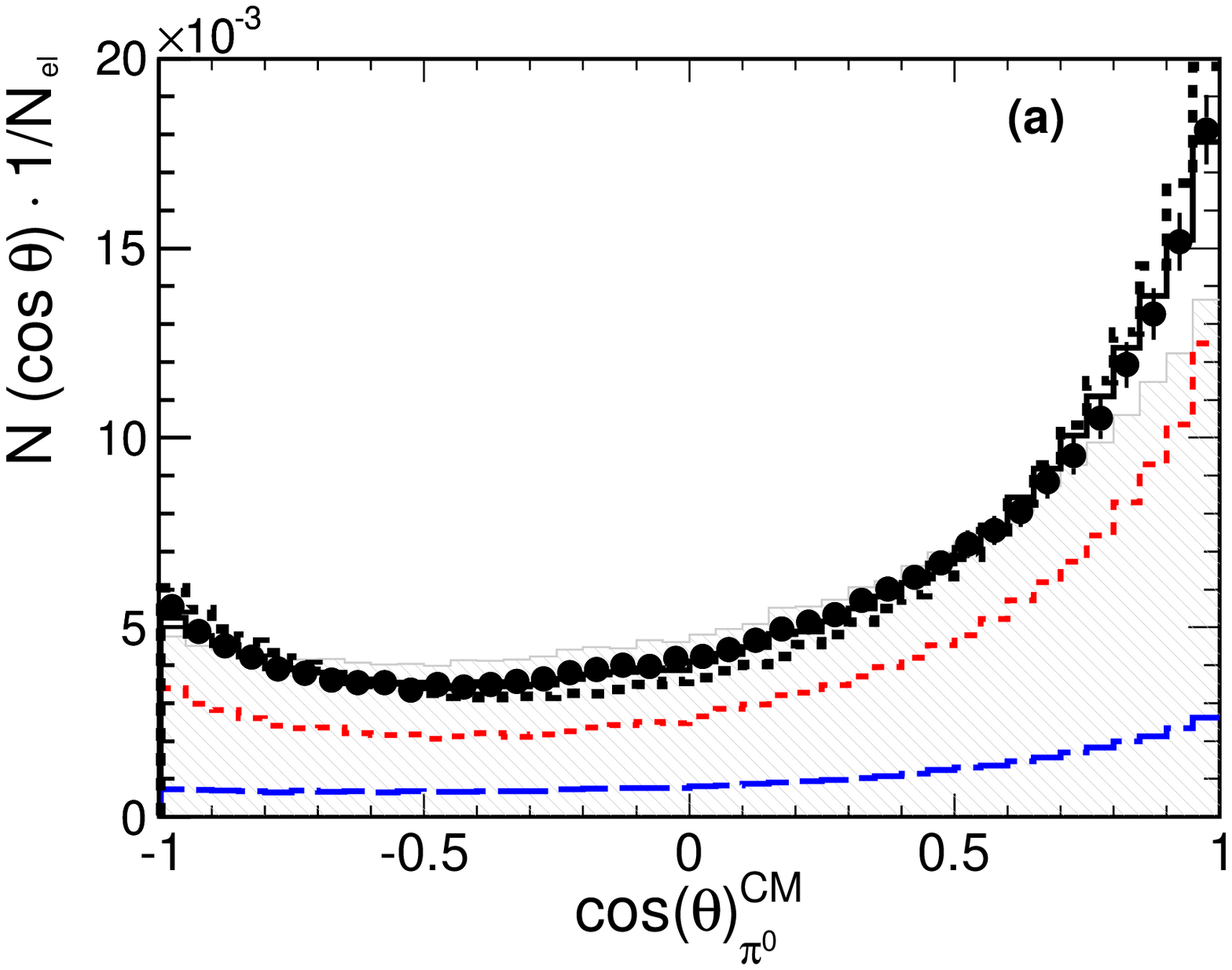}
                    \includegraphics[width=0.24\textwidth]{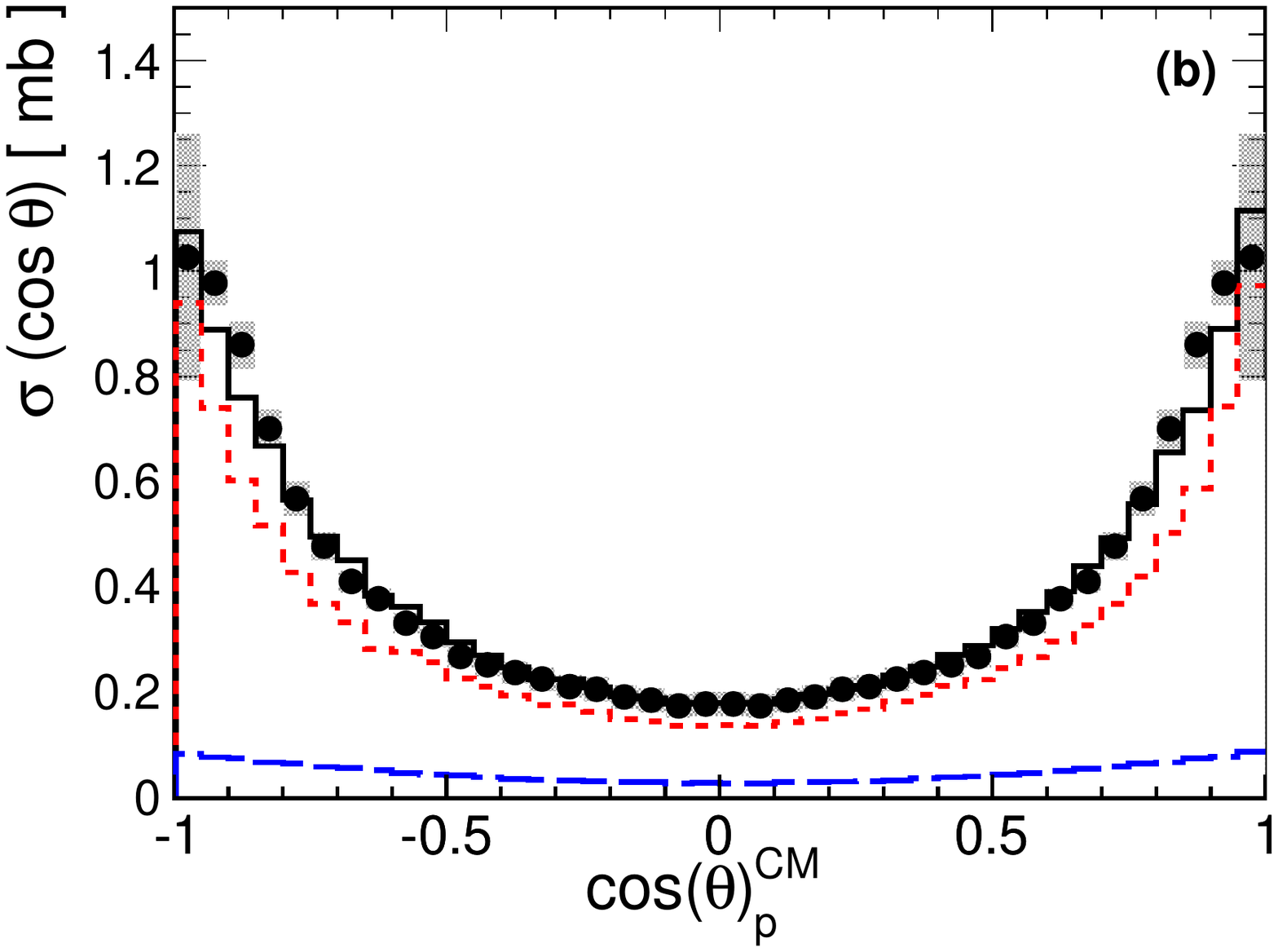}
                    \includegraphics[width=0.24\textwidth]{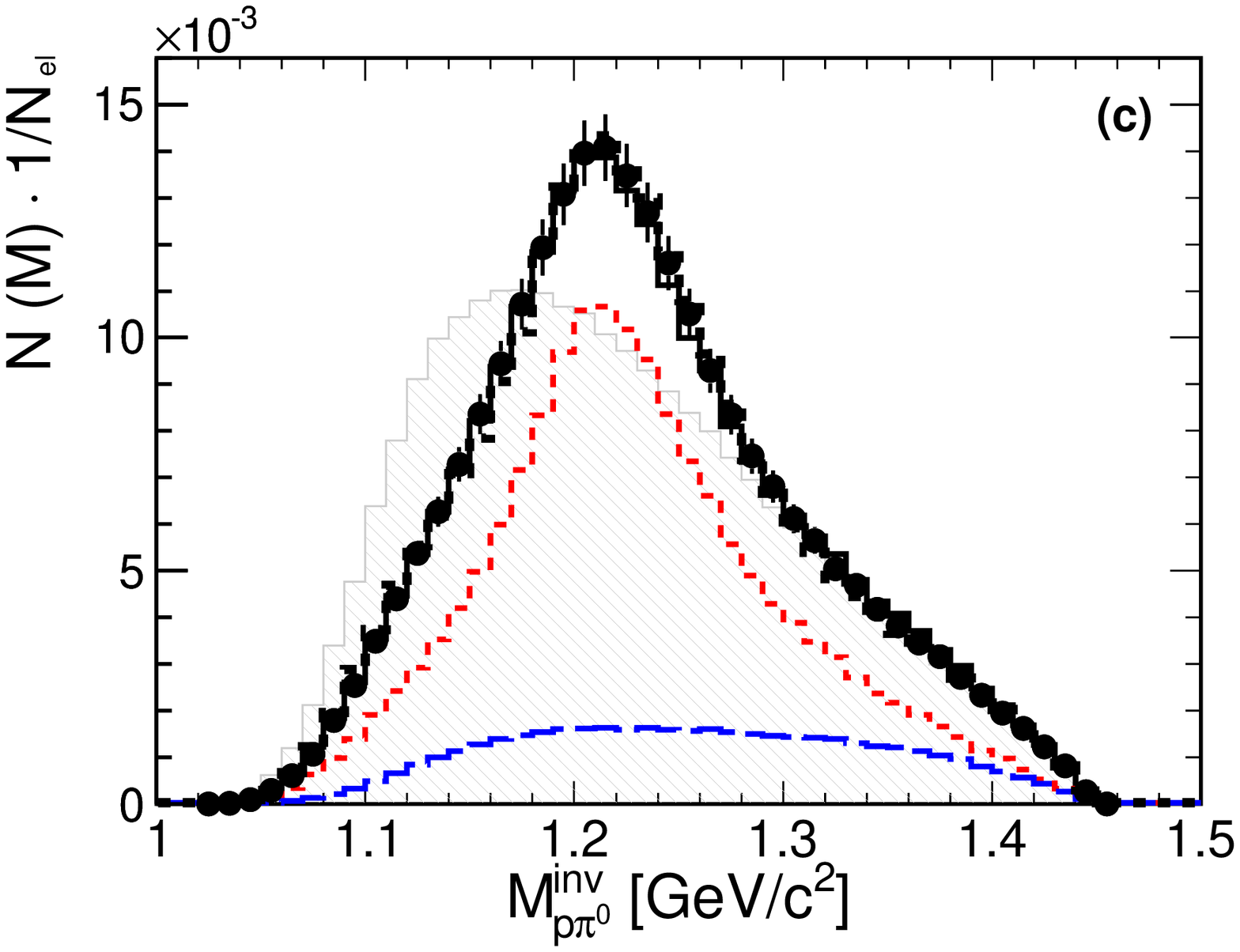}
                    \includegraphics[width=0.24\textwidth]{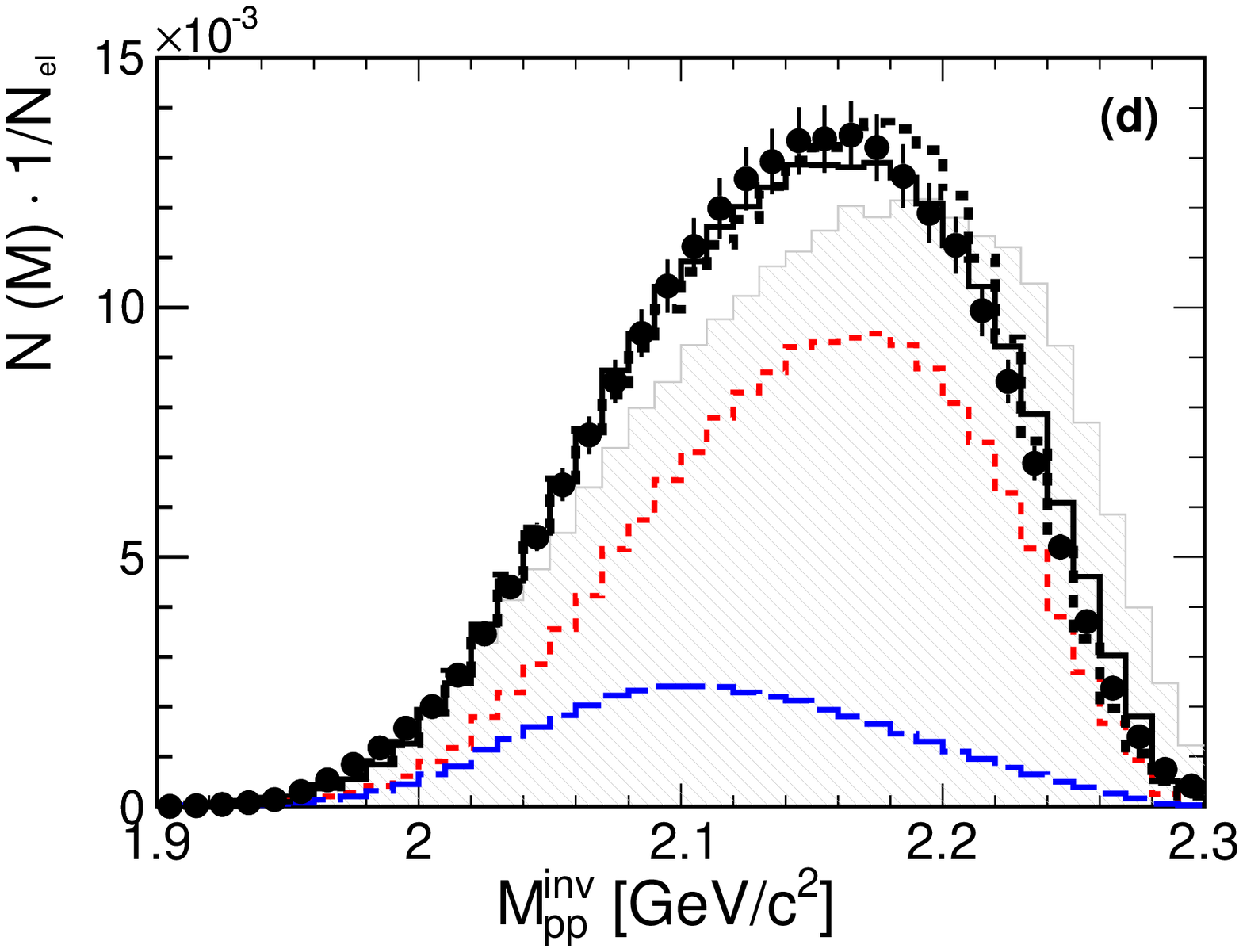}\\
                    Angular distribution of (a) $\pi^{0}$ and (b) $p$ in c.m.s. reference frame. Invariant mass of (c) $p\pi^{0}$ and (d) $pp$.\\

                    \includegraphics[width=0.24\textwidth]{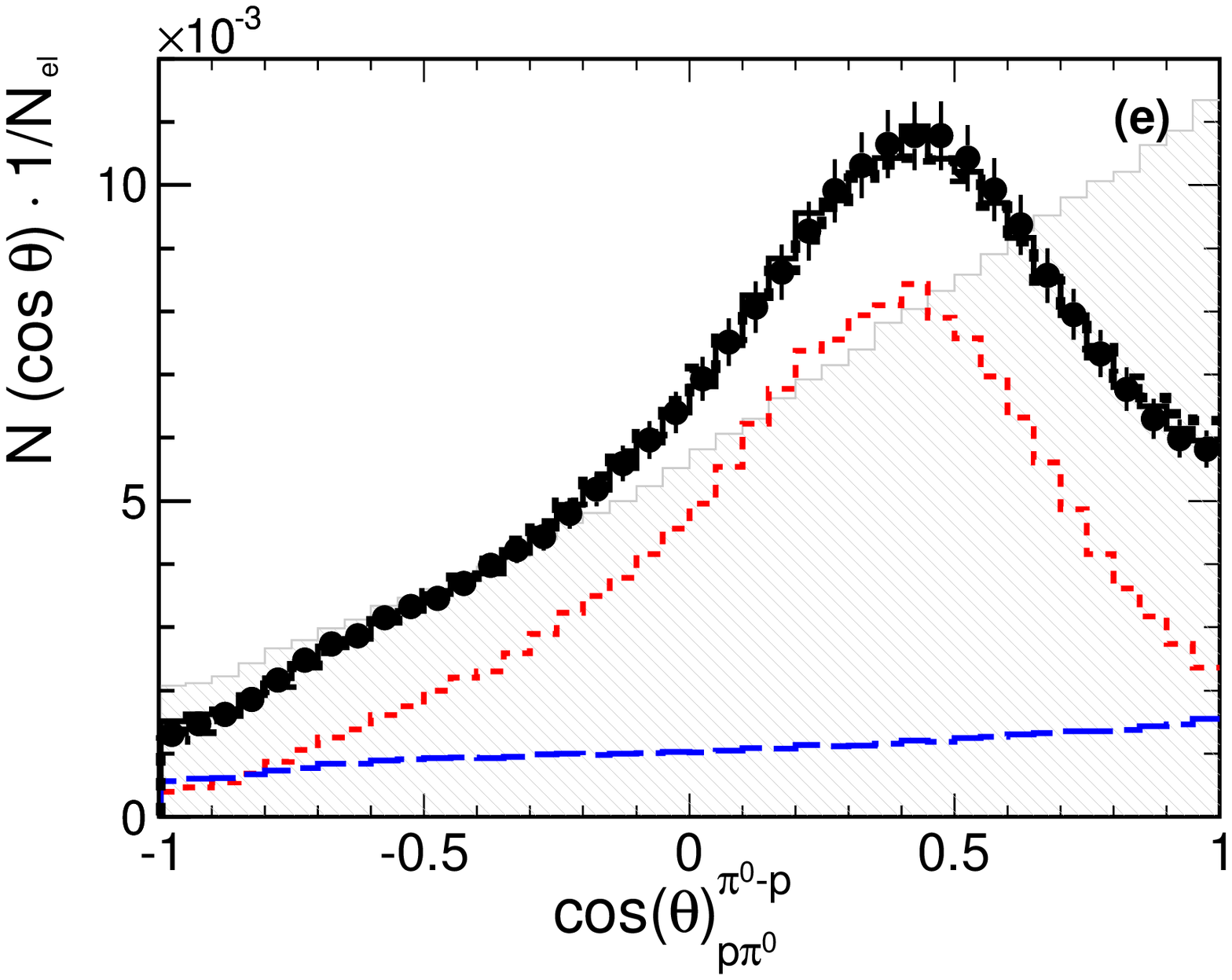}
                    \includegraphics[width=0.24\textwidth]{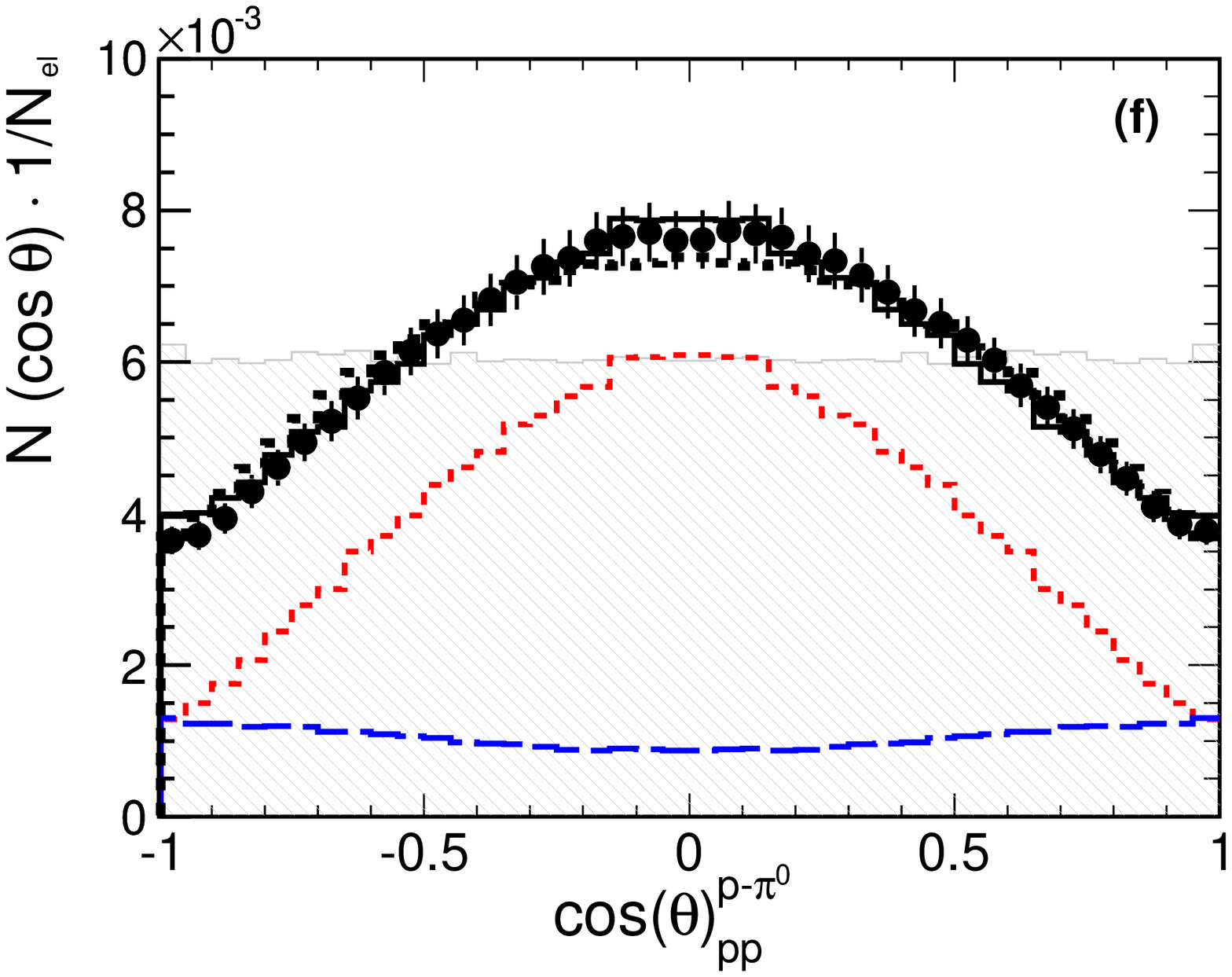}
                    \includegraphics[width=0.24\textwidth]{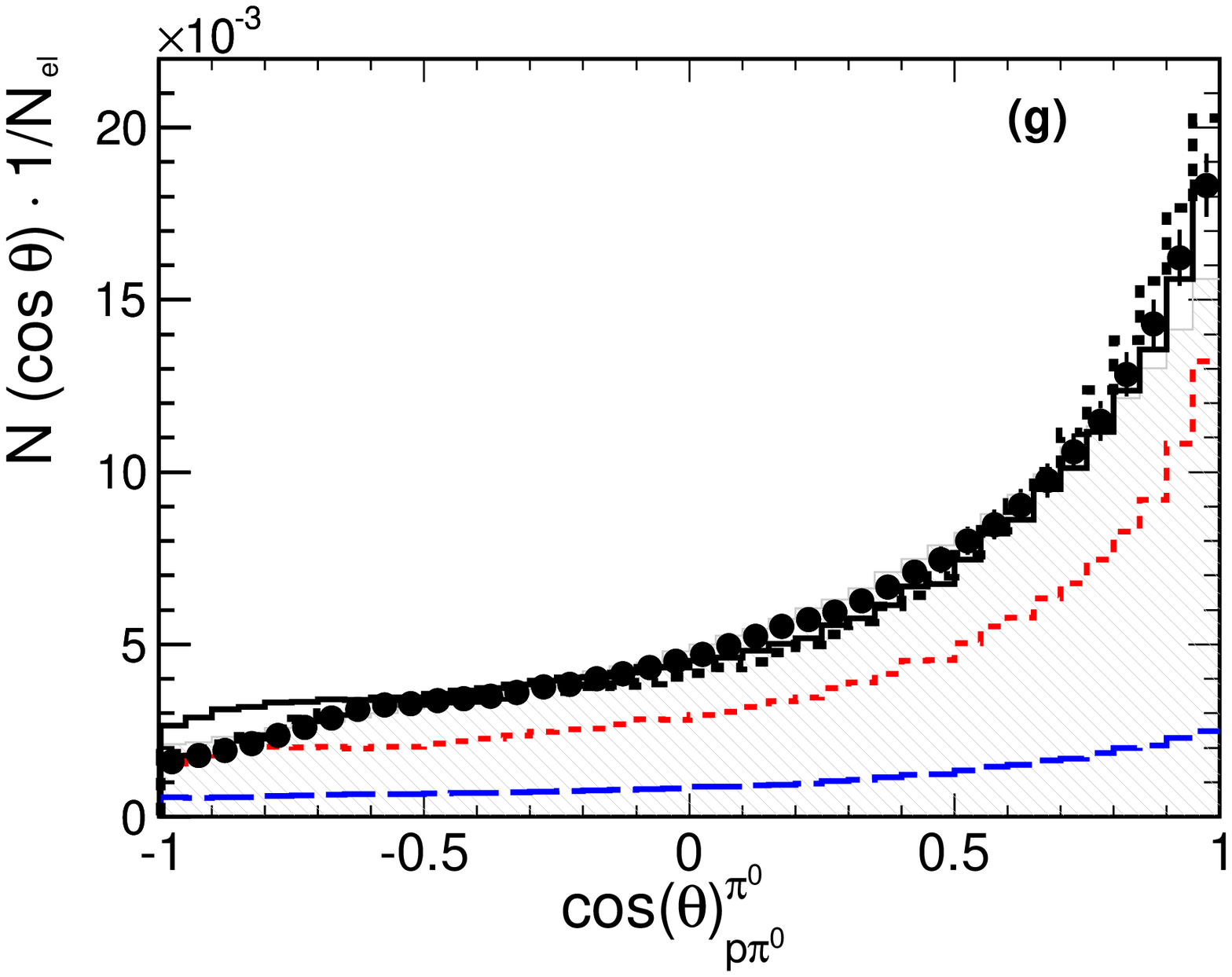}
                    \includegraphics[width=0.24\textwidth]{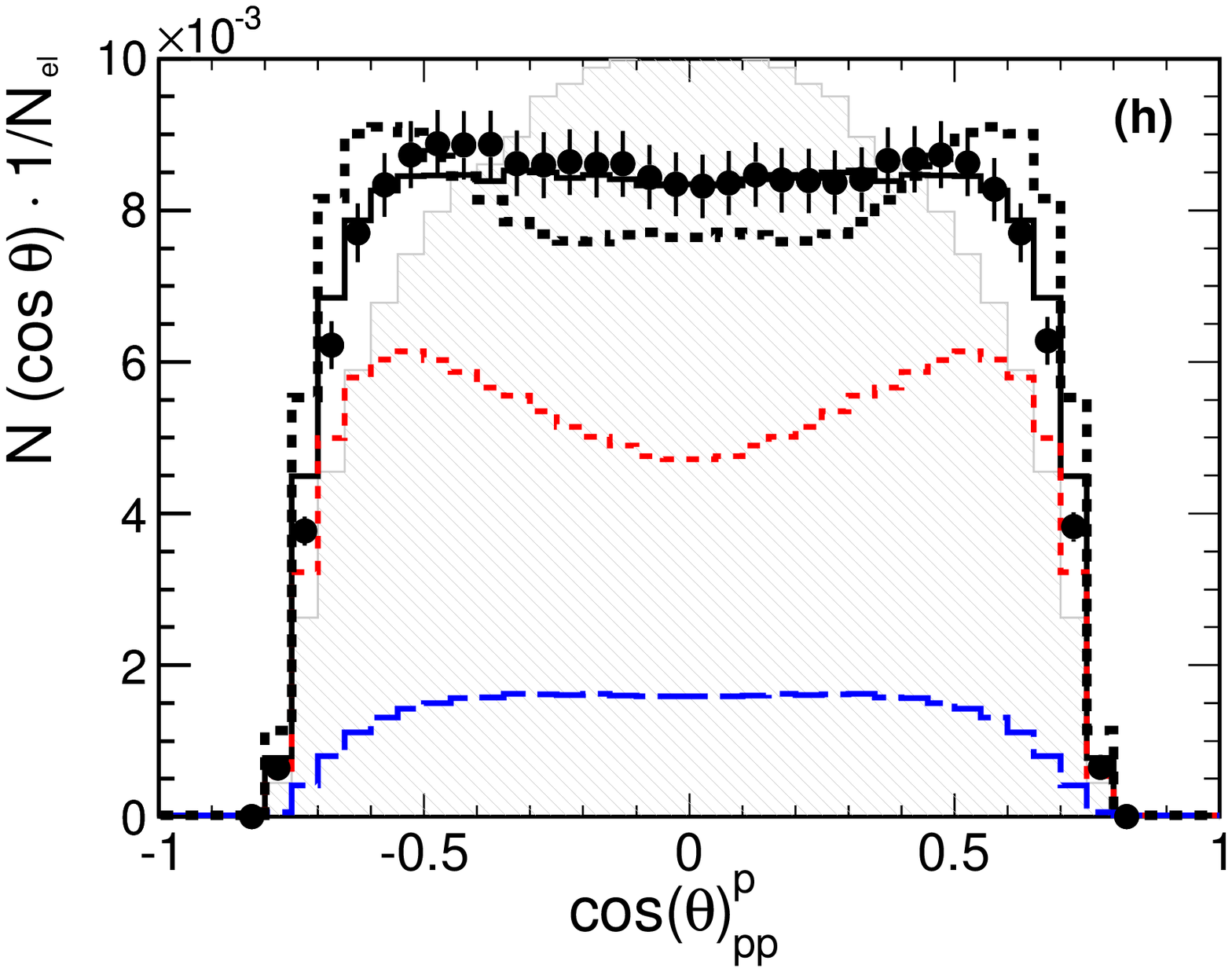}\\
                    Helicity distribution of (e) $\pi^{0}$ in $p\pi^{0}$ reference frame and (f) $p$ in $pp$ reference frame. Angular distribution of (g) $\pi^{0}$ in $p\pi^{0}$ GJ reference frame and (h) $p$ in $pp$ GJ reference frame.\\
            \caption{(Color online) Various projections for the $pp\pi^{0}$ channel. Uncorrected data points (black) within the HADES acceptance with systematic and statistical error bars, normalized to the number of $pp$ elastic scattering ($N_{el}$). Histograms: total PWA solution (solid black), the $\Delta$(1232) contribution (short-dashed red) and the $N$(1440) contribution (long-dashed blue). Dotted histogram (black): modified resonance model. The grey hatched area in each panel shows the distribution in the case of isotropically simulated particles.}
    \label{f2}
    \end{figure*}

\begin{figure*}
            \centering
                    \includegraphics[width=0.3\textwidth]{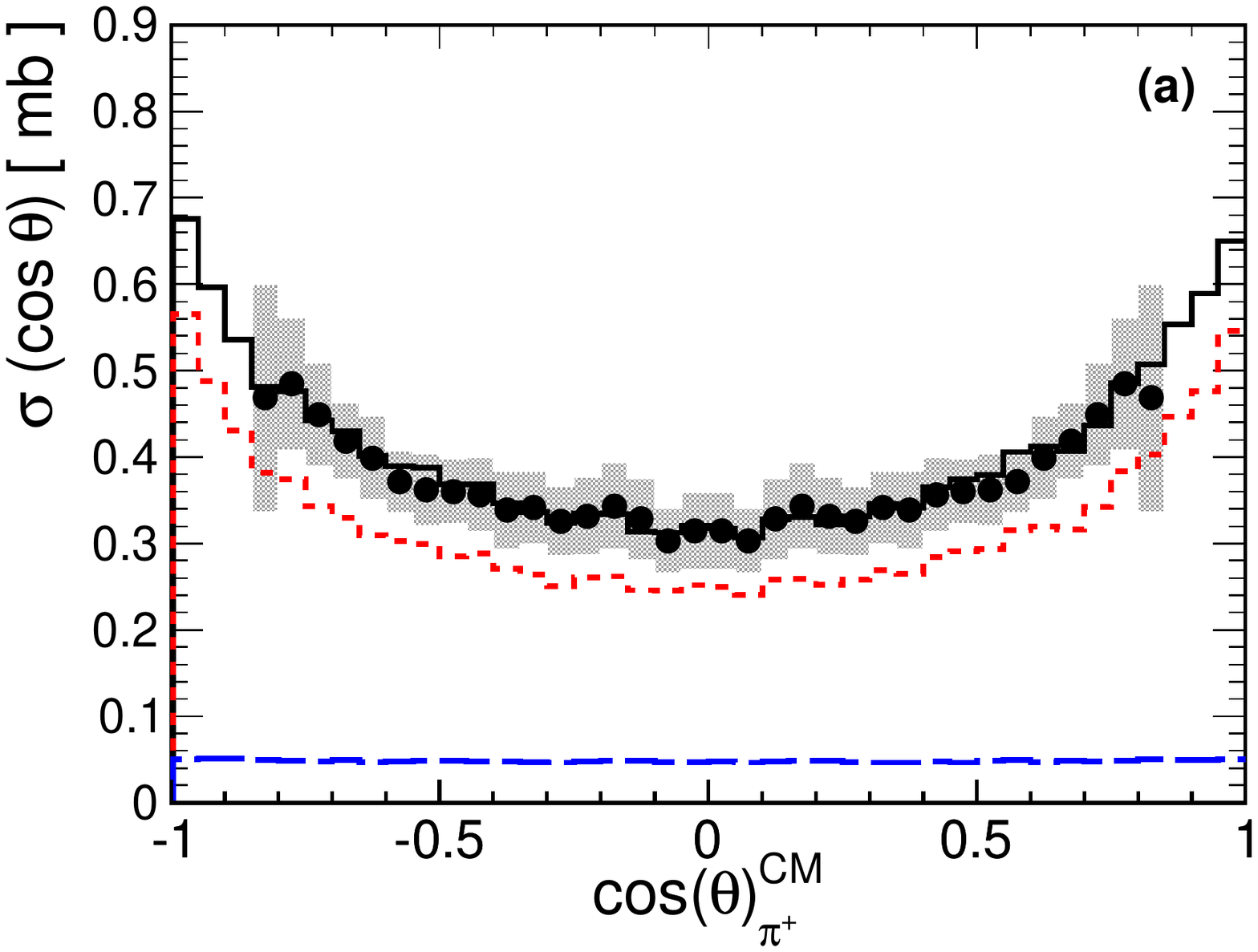}
                    \includegraphics[width=0.3\textwidth]{p_costh}
                    \includegraphics[width=0.3\textwidth]{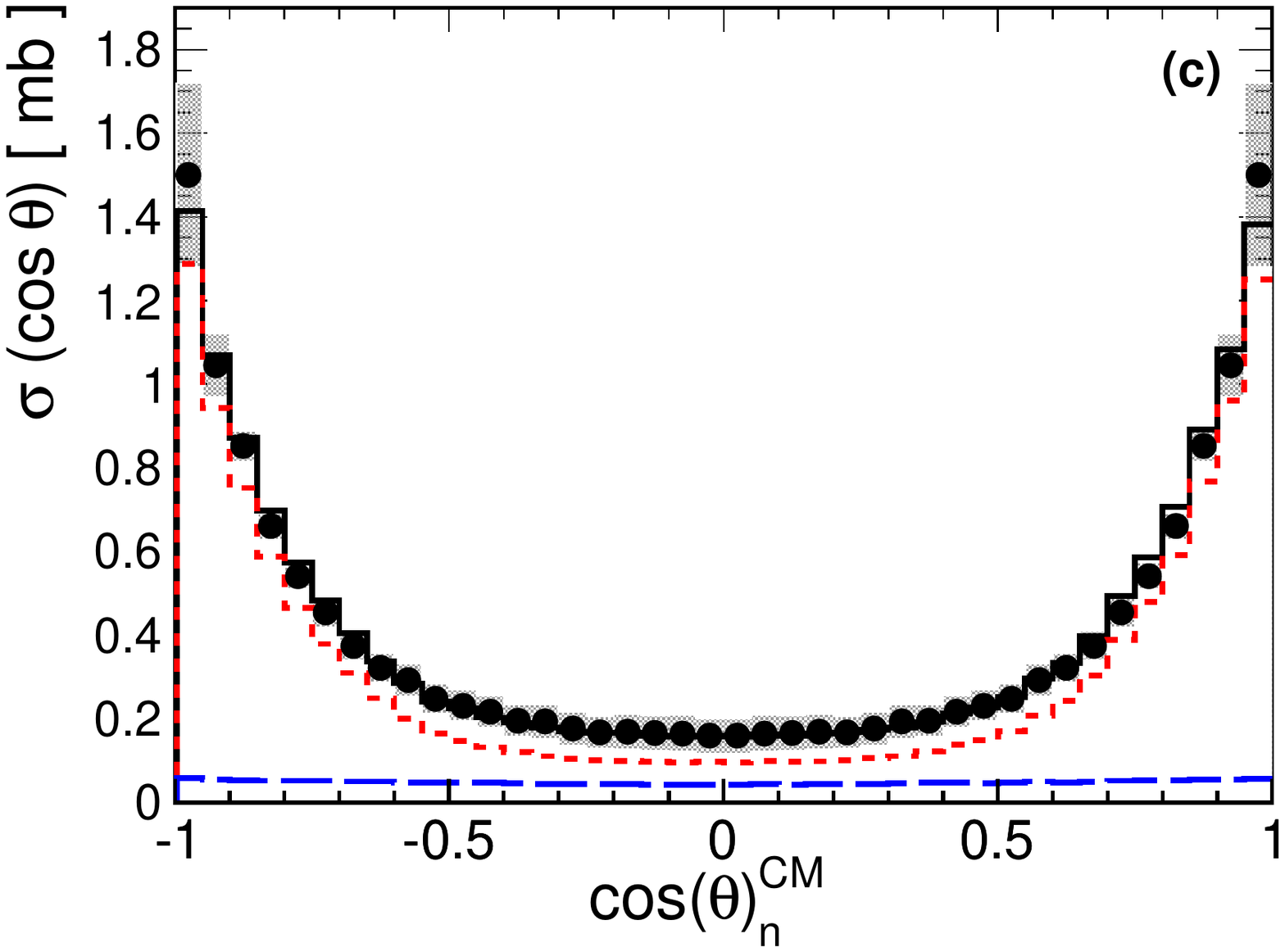}\\
                    Acceptance and efficiency corrected angular distribution of (a) $\pi^{+}$, (b) $p$ and (c) $n$ in c.m.s. reference frame.\\

                    \includegraphics[width=0.3\textwidth]{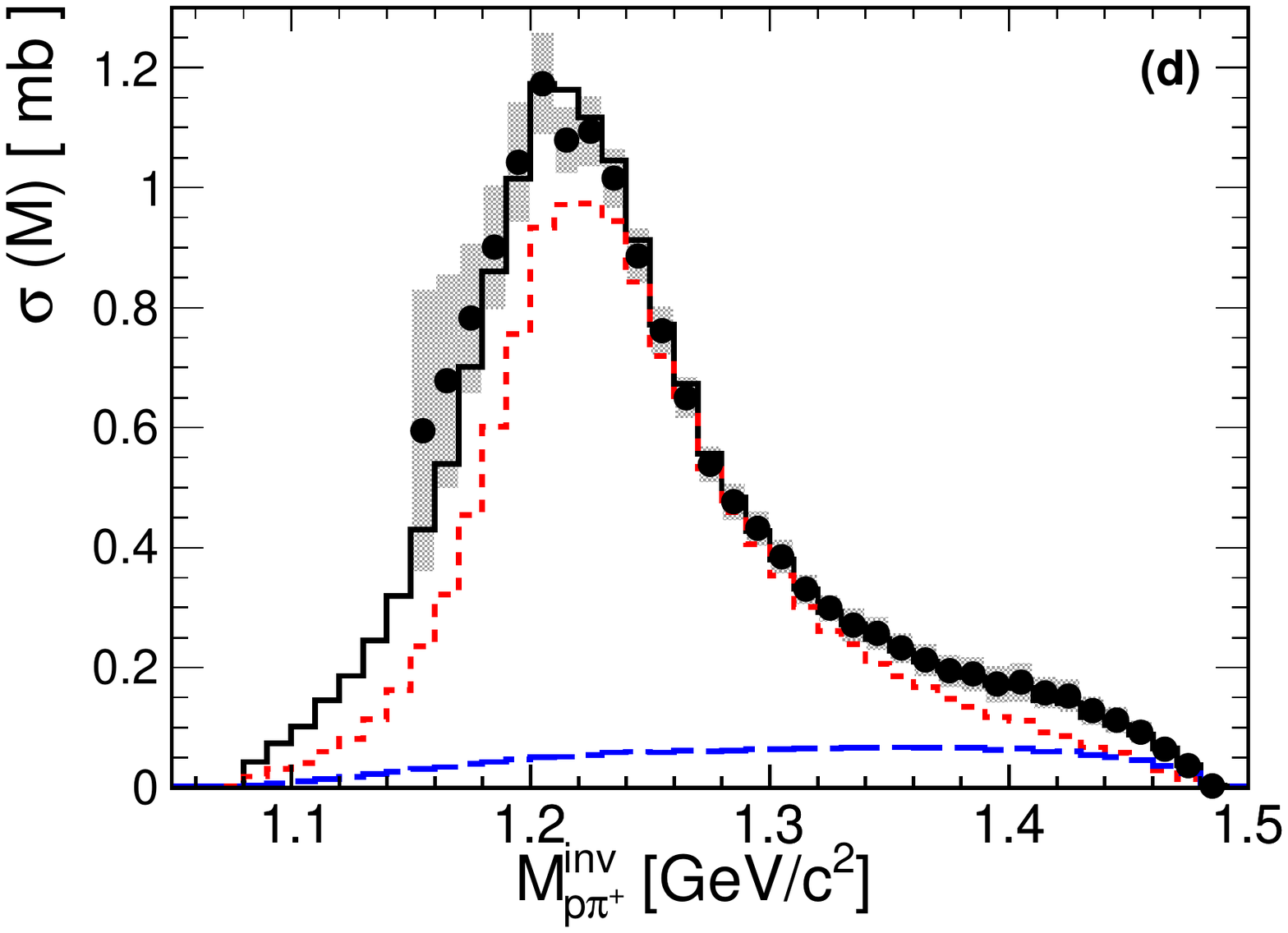}
                    \includegraphics[width=0.3\textwidth]{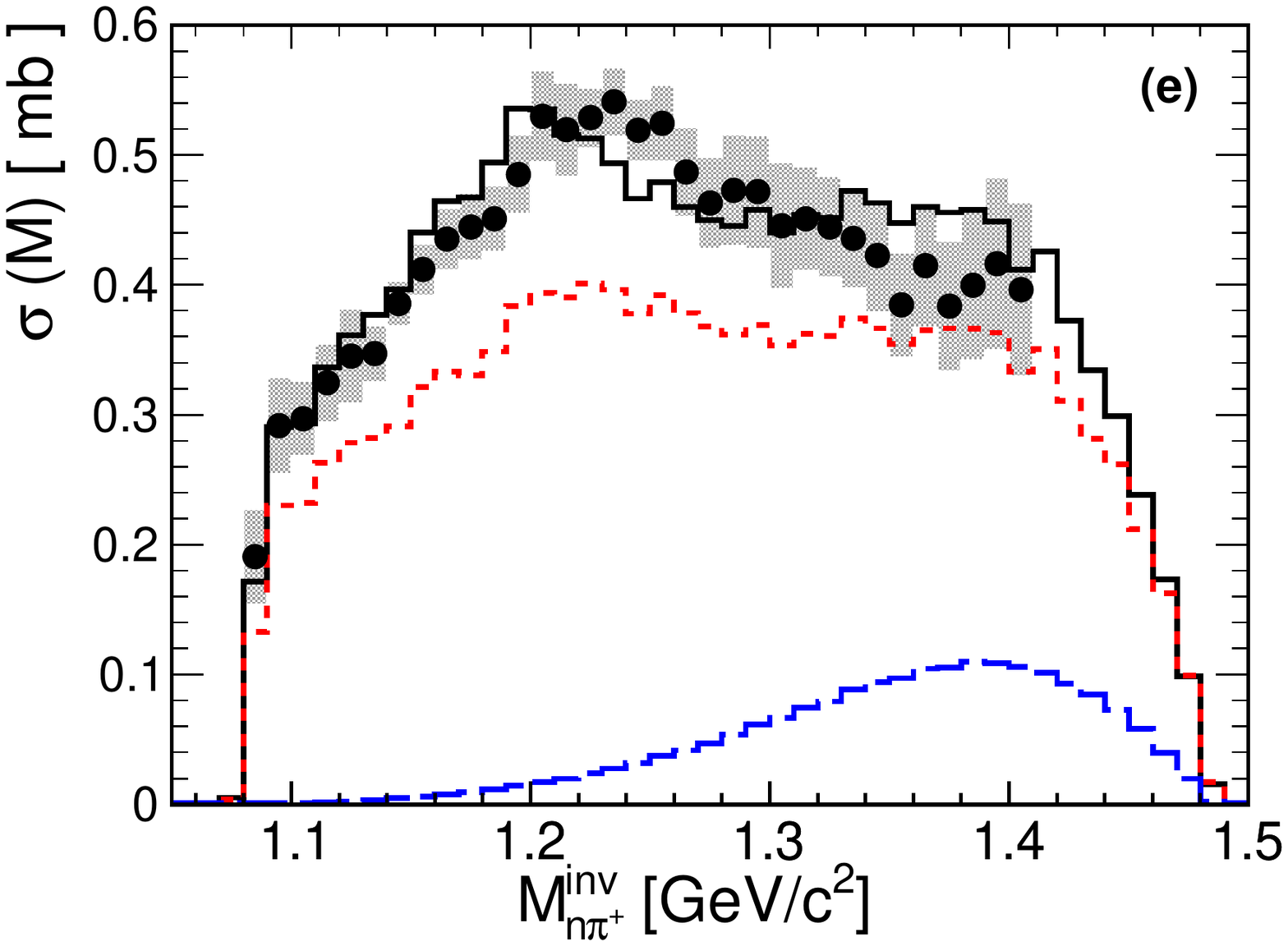}
                    \includegraphics[width=0.3\textwidth]{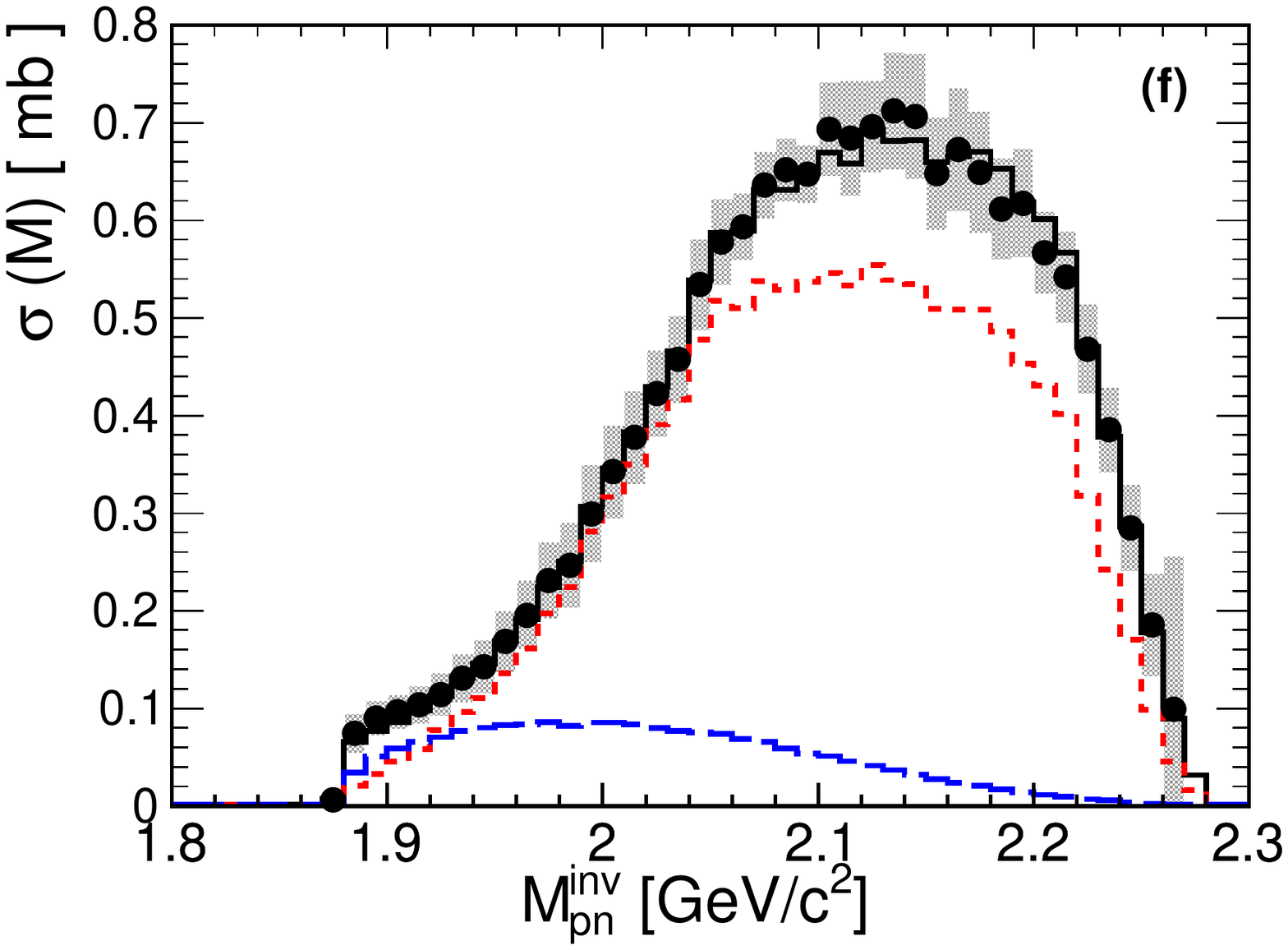}\\
                    Acceptance and efficiency corrected invariant mass of (d) $p\pi^{+}$, (e) $n\pi^{+}$ and (f) $pn$.\\

                    \includegraphics[width=0.3\textwidth]{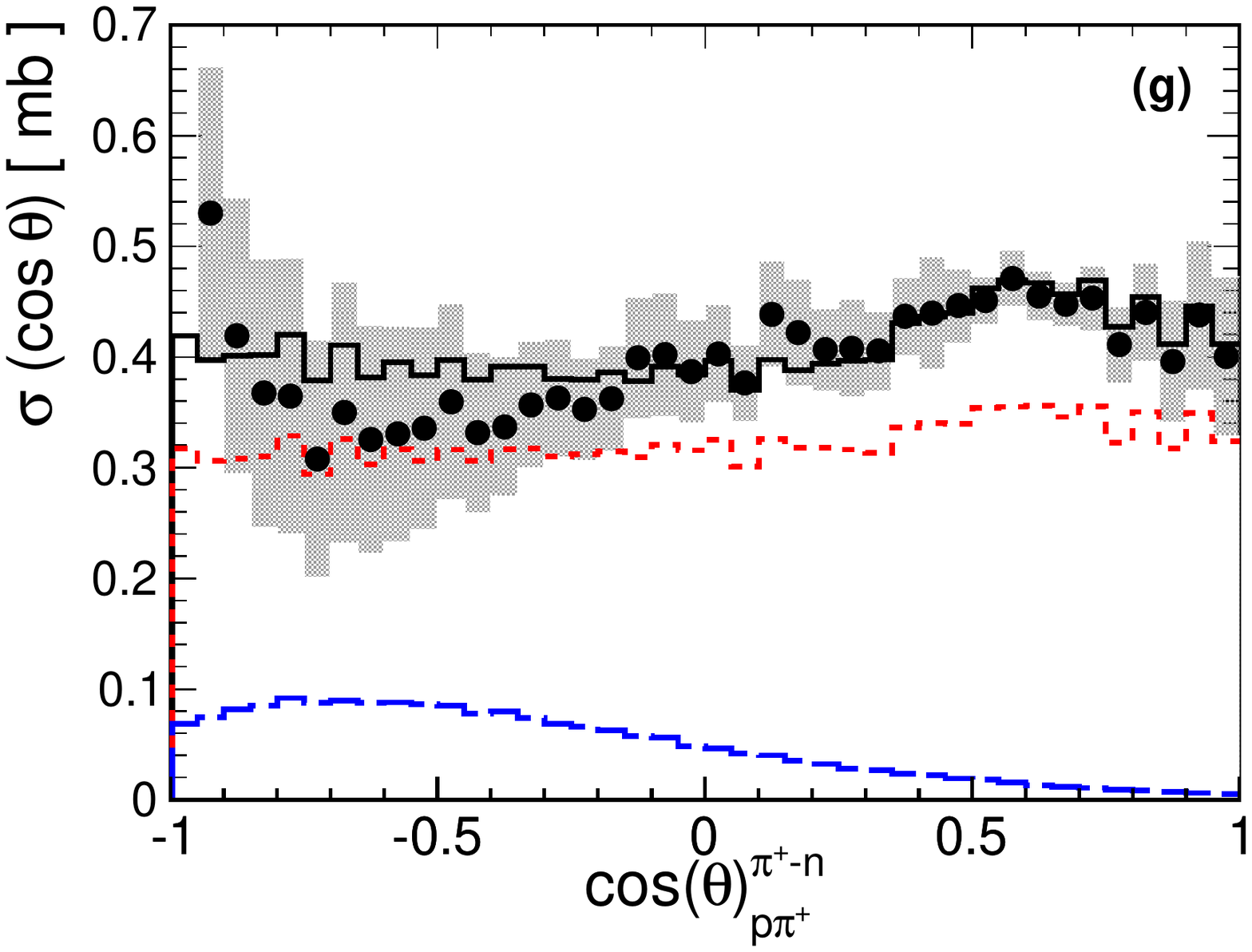}
                    \includegraphics[width=0.3\textwidth]{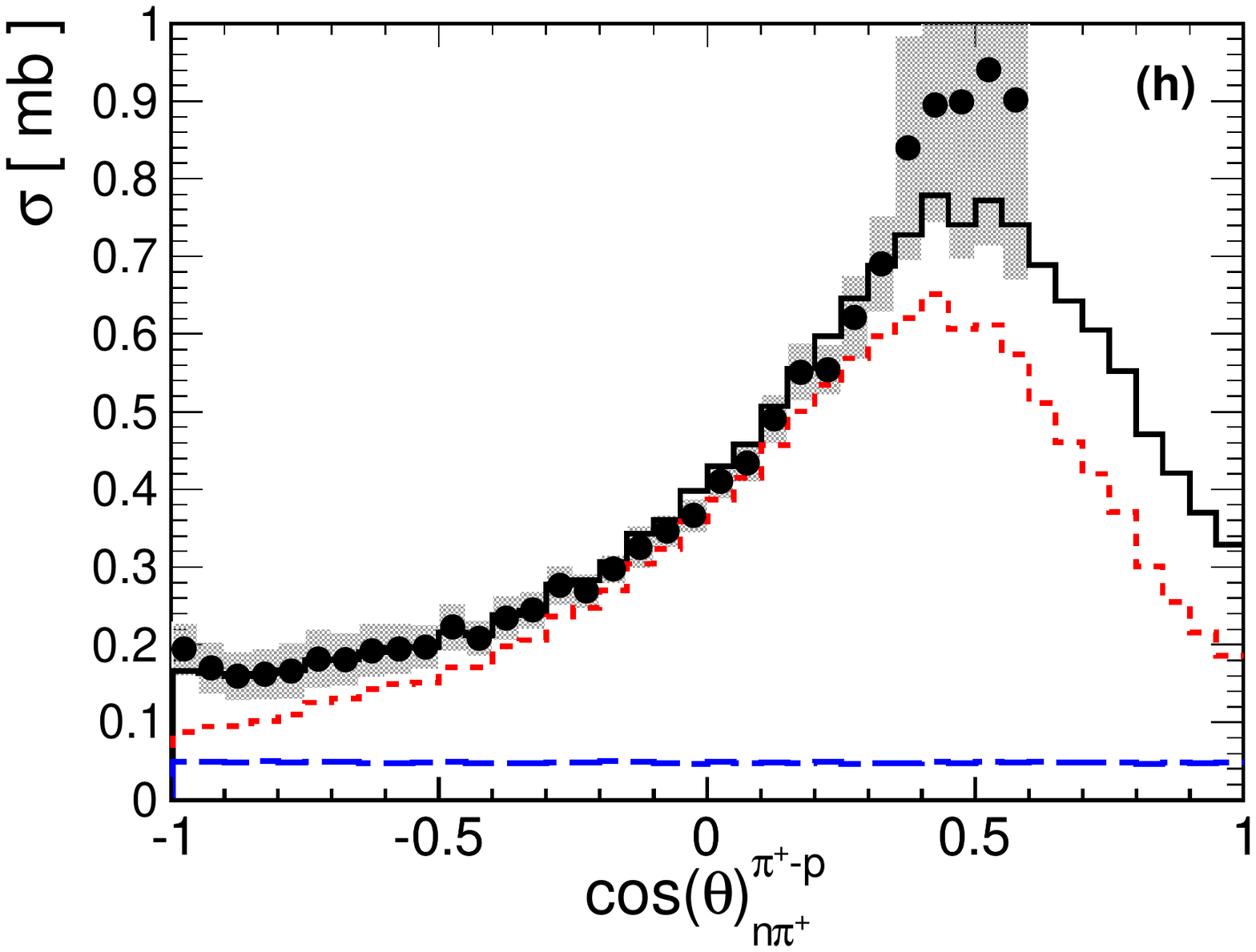}
                    \includegraphics[width=0.3\textwidth]{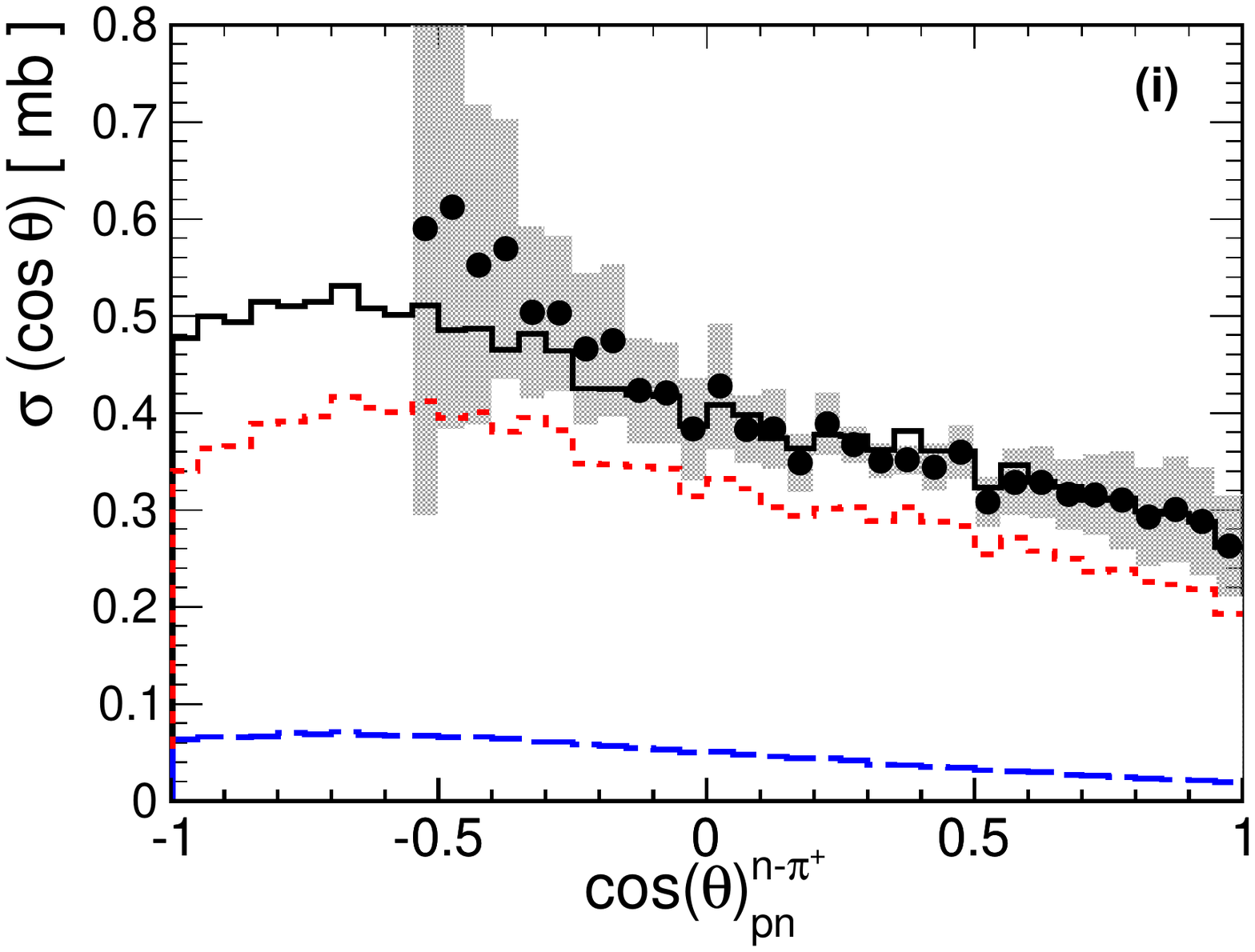}\\
                    Acceptance and efficiency corrected helicity distribution of (g) $\pi^{+}$ in $p\pi^{+}$ reference frame, (h) $\pi^{+}$ in $n\pi^{+}$ reference frame and (i) $n$ in $pn$ reference frame.\\
            
                    \includegraphics[width=0.3\textwidth]{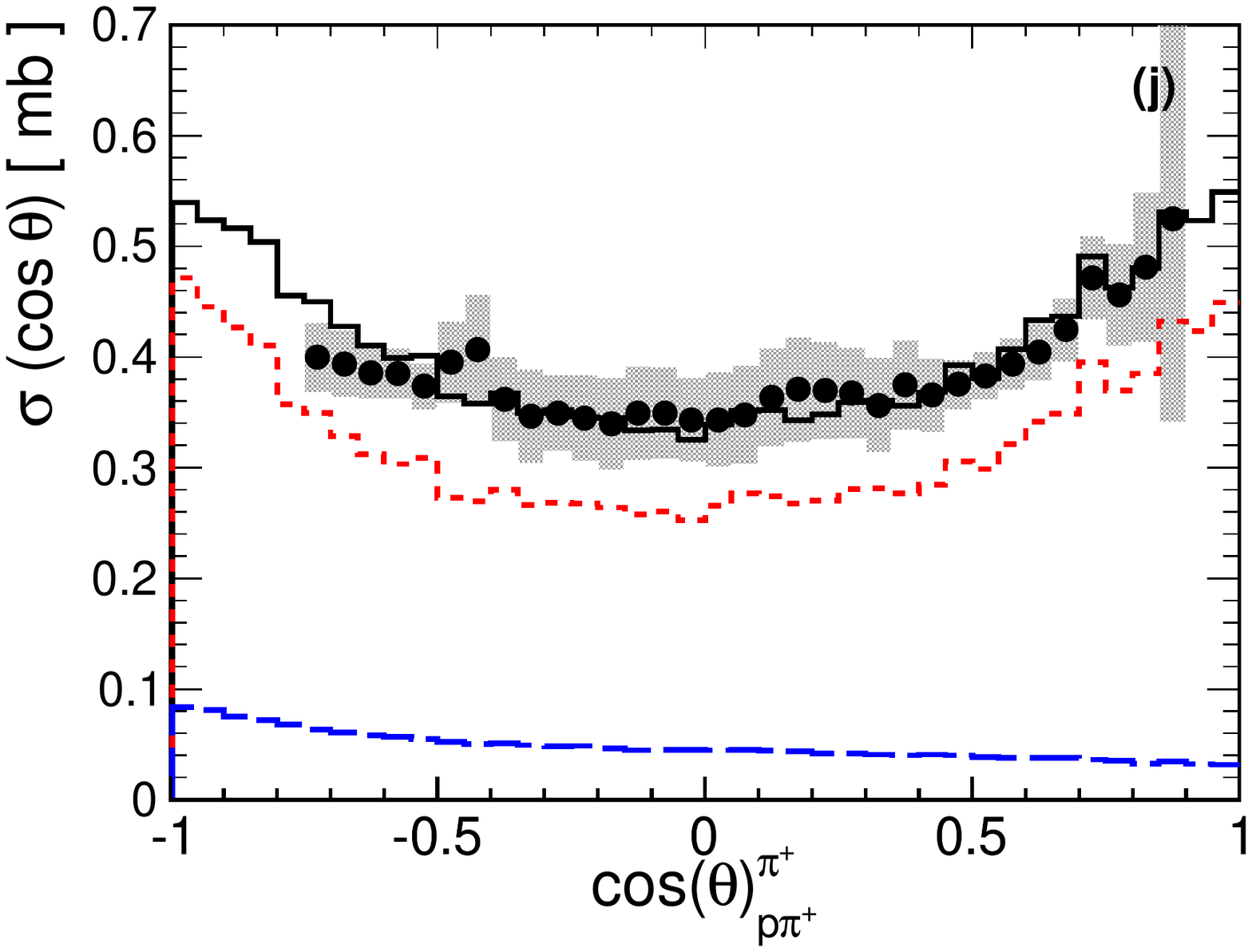}
                    \includegraphics[width=0.3\textwidth]{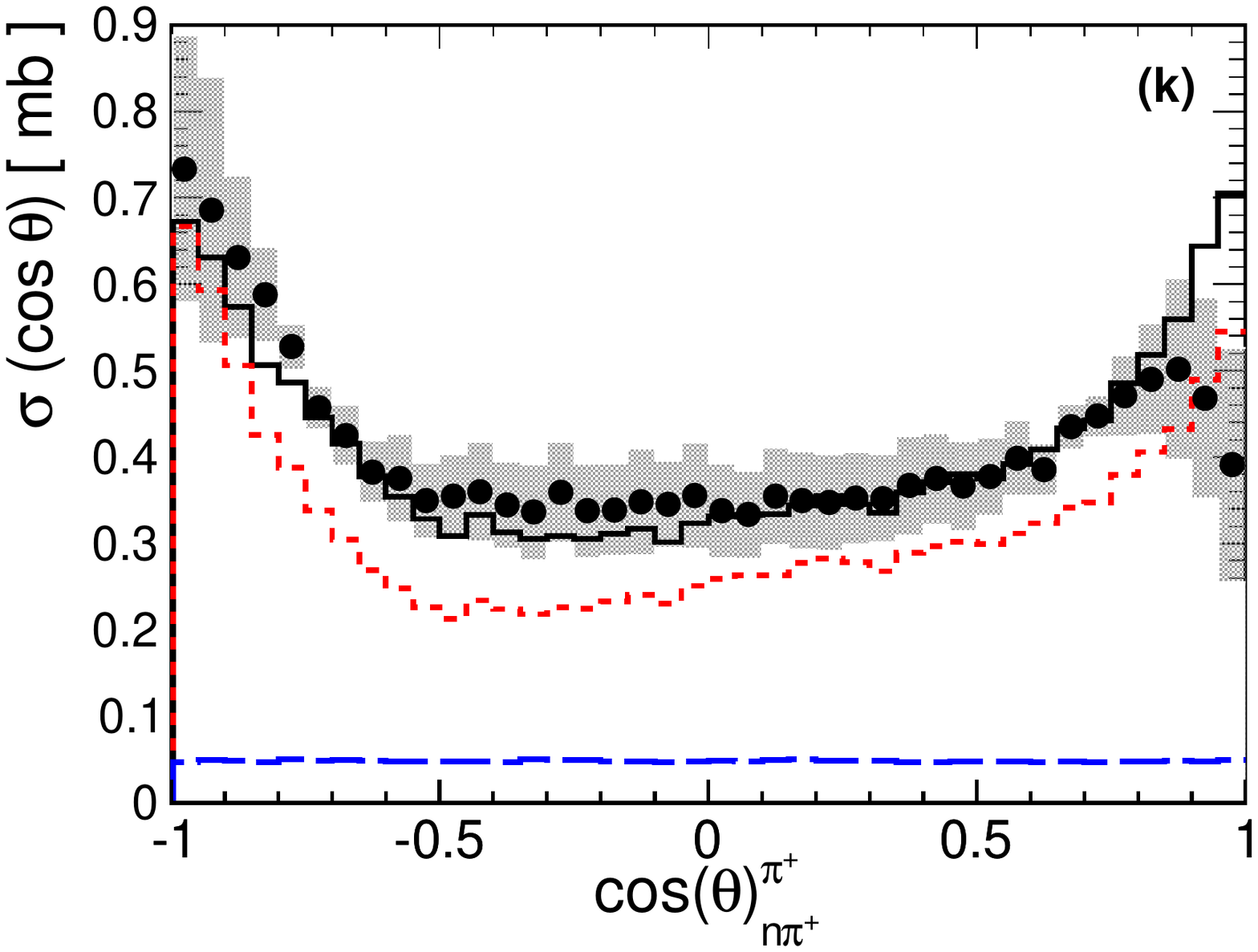}
                    \includegraphics[width=0.3\textwidth]{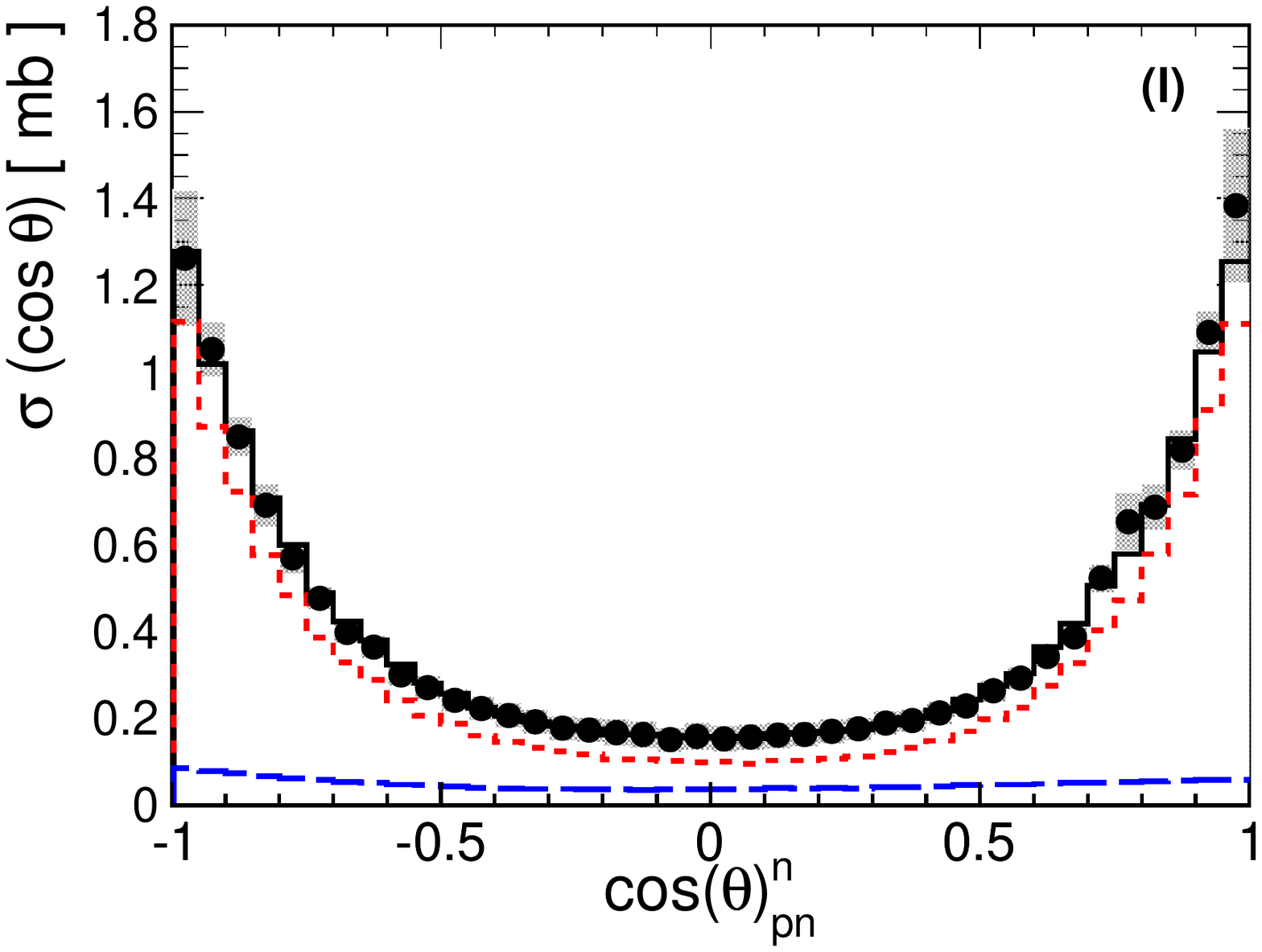}\\
                    Acceptance and efficiency corrected angular distribution of (j) $\pi^{+}$ in $p\pi^{+}$ GJ reference frame, (k) $\pi^{+}$ in $n\pi^{+}$ GJ reference frame and (l) $n$ in $pn$ GJ reference frame.\\

            \caption{(Color online) $np\pi^{+}$ channel: data points after acceptance corrections (black dots) based on the partial wave analysis solution. Data points in the areas of very low acceptance are omitted. Uncertainties originating from the various PWA solutions (as explained in the text) and statistical errors are visualized as grey band. Normalization error is not indicated. Histograms: total PWA solution (solid black), the $\Delta$(1232) contribution (short-dashed red) and the $N$(1440) contribution (long-dashed blue).}

    \label{f3}
    \end{figure*}

\subsection{\bf{$pp\pi^{0}$} channel}
\label{pppi0_channel}

The identification of two protons in the HADES spectrometer results in a reduced acceptance for the $pp\pi^0$ reaction channel. As pointed out in \cite{Ramstein2012}, the resonance model \cite{Teis1997} does not reproduce satisfactorily our measured observables in the $pp\pi^{0}$ channel. To improve the description, the aforementioned parameterization of the resonance angular distribution, deduced from the $np\pi^{+}$ channel  analysis, was applied for the $\Delta^{+}$ production. Although some angular projections still unravel slight discrepancies between the data and the model, the overall description is quite good (see Fig. \ref{f2}) and allows for the correction of the data for the reconstruction inefficiencies and detector acceptance, with the model-driven extrapolation, in an analogous way as it was done for the $np\pi^{+}$ channel. The integrated correction factor in the $pp\pi^{0}$ channel varies in the range 15-25, depending on the distribution. The deduced total cross section amounts to 3.87~\er~0.60 mb. Due to the smaller, as compared to the $np\pi^{+}$ channel, acceptance coverage, the systematic error related to the acceptance corrections is $12\%$ (estimated as in the previous case), the background subtraction error is similar and amounts to $6\%$.

Taking into account the isotopic relations between the final state channels, one gets the total cross section for the $\Delta^{+}$ production equal to either $4.98\pm0.72$ mb (deduced from the $np\pi^{+}$ channel, where the $\Delta^{++}$ contribution is $14.86\pm2.19$ mb) or $5.42\pm0.69$ mb (deduced from the $pp\pi^{0}$ channel). The expected ratio $\sigma_{\Delta^{++}}/\sigma_{\Delta^{+}}$ is 3, which is satisfied within the errors in both cases: $2.98\pm 0.61$ or $2.74\pm 0.53$, respectively. One should notice that the $N$(1440) contribution is negligible in the resonance model approach.

\section{Partial wave analysis results}
\label{PWA_results}

The above modified resonance model describes fairly well the angular and mass distributions and can be used for the acceptance correction of the data. However, the introduced modification of the angular distributions of the $\Delta$ resonance does not provide insight into the production mechanism. Moreover, the non-resonant contribution is completely neglected and $N$(1440) contribution is treated in a very simplified manner.

The successful analysis in the framework of the Bonn-Gatchina PWA was already demonstrated in the case of $p+p$ data measured at PNPI at lower energies (see \cite{Ermakov2014} and \cite{Ermakov2011}) and for the pK$\Lambda$ final state \cite{Agakishiev2015} in the case of $p+p$ data at a kinetic beam energy of 3.5 $GeV$ measured with HADES. In this approach, the total reaction amplitude $A$ is described as a sum of partial wave amplitudes with the corresponding angular dependencies:
\be
A=\sum\limits_\alpha A^\alpha_{tr}(s) Q^{in}_{\mu_1\ldots\mu_J}(S
LJ)A_{2b}(j,S_2L_2 J_2)(s_j)\times
\nonumber\\
Q^{fin}_{\mu_1\ldots\mu_J}(j,S_2L_2J_2S'L'J)\ .
\label{ampl}
\ee
Here $S$, $L$ and $J$ are the spin, the orbital momentum and the total angular momentum of the initial $NN$ system, $S_2$, $L_2$ and $J_2$ denote spin, orbital momentum and total angular momentum of the two-particle system in the final state, and $S'$ and $L'$ are spin and orbital momentum between this two-particle system and the spectator particle with index $j$, e.g. $\pi$(1), $p$(2), $n$(3). The invariant mass of the two-body system is determined by $s_j=(P-q_j)^2$, where $q_j$ is the four-momentum of the spectator and $P$ is the total momentum of the reaction. The operators $Q^{in}$ and $Q^{fin}$ are tensors of the rank $J$ constructed for each event from the momenta of the initial and final state particles. Their convolution provides the angular dependence of the amplitude; the explicit form is given in \cite{Anisovich2007}. For the transition amplitude $A_{tr}^\alpha$ from the initial $NN$ to the $NN\pi$ system, we introduced the multi-index $\alpha$ that summarizes all quantum numbers described above. The differential cross section calculated from this amplitude is maximized for the data events with the event-by-event maximum likelihood method, thus taking into account all correlations in the multidimensional phase space.

The likelihood function is normalized by the Monte Carlo integral calculated with events generated according to the phase space distribution passed through the simulated detector response and signal reconstruction. It means that the distribution of these events weighted by the cross section from the found solution should closely reproduce, within the HADES acceptance, the distribution of the experimental data. The solution provides also a possibility to extrapolate the cross section to the region with low experimental efficiency and therefore to perform the acceptance correction of the data.

The resonance production in the $\pi$N channel is parameterized by relativistic Breit-Wigner amplitudes. For the $\Delta$ and Roper states we introduce the following parameterization ($j$=2, 3):
\be
A_{2b}(j,\beta)(s_{\pi N})= \frac{g^R_{\pi N}}{M_R^2-s_{\pi N}-i
M_R\Gamma_R}\,\qquad,
\ee
where the multi-index $\beta$ stands for $S_{\pi N},L_{\pi N},J_{\pi N}$. The resonance total width is equal to the sum of partial widths, and the $g^R_{\pi N}$ coupling is connected with the $\pi$N partial width by:
\be
M_R \Gamma_{\pi N}=(g^R_{\pi N})^2\frac{2k_{\pi N}}{\sqrt s_{\pi N}}\frac{1}{16\pi}\frac{k^{2L}_{\pi N}}{F(k^2_{\pi N},L_{\pi N},r)}\,.
\label{pwidth}
\ee
Here, the quantity $k_{\pi N}$ is the relative momentum of the pion and nucleon in the $\pi$N rest frame, and $F(k^2_{\pi N},L_{\pi N},r)$ denotes the Blatt-Weiskopf form factor with interaction radius $r$ \cite{Anisovich2006}. 

Equation \ref{pwidth} defines the energy dependence of the resonance partial width. The initial values of masses and total widths of the resonances were taken from the review of the Particle Data Group \cite{PDG2014} and adjusted in the course of the fit procedure. The interaction radius $r$ was fixed at 0.8 fm. The total width of the $\Delta$ state is completely defined by the decay into the $\pi$N system with $L_{\pi N}$=1 ($S_{\pi N}$=$\frac 12$, $J_{\pi N}$=$\frac 32$). This form of Blatt-Weiskopf parameterization is also used in the Manley and Saleski partial wave analysis fit \cite{Manley1992}. The difference of the cut-off function, as compared to the Moniz parameterization Eq. (\ref{e_MONIZ}), is not so pronounced at the energy of 1.25 GeV, but becomes important at higher energies \cite{Hades2014}. 
In the case of the Roper resonance, the $\pi$N partial width contributes about 65\% to the total width of the state. In general, the partial widths defined by the two pion-nucleon channel should have a complicated energy dependence. Possible parameterizations of the Roper resonance do not change the solutions very much, as to be discussed below.

The non-resonant contributions in the $NN$ scattering channel are parameterized by a modified scattering length approximation expression ($j$=1):
\begin{equation}
A_{2b}(j,\beta)(s_{NN})=\frac{r_\beta a_\beta\sqrt{s_{NN}}}
{1\!-\!\frac 12 r_\beta k_{NN}^{2}a_\beta\!+\! \frac{ ia_\beta k_{NN}^{2L_\beta+1} }{ F(k^2_{NN},r_\beta,L_{NN}) }},
\label{a_NN}
\end{equation}
where $k_{NN}$ is the nucleon-nucleon relative momentum calculated in the $NN$ rest system, $L_{NN}$ is the orbital momentum of the $NN$ system, $r_\beta$ is the effective range and $a_\beta$ is the scattering length of the system ($\beta=S_{NN},L_{NN},J_{NN}$). For the $S$-waves, Eq. (\ref{a_NN}) corresponds to the scattering-length approximation formula suggested in \cite{Watson1952,Migdal1955}. The pn scattering length and effective range are fixed for the $S$-waves at $a(^{2S+1}L_J)=a(^1S_0)=-23.7$ fm, $r(^1S_0)=2.8$ fm and $a(^3S_1)=5.3$ fm and $r(^3S_1)=1.8$ fm.

\begin{figure*}
            \centering
                    \includegraphics[width=0.24\textwidth]{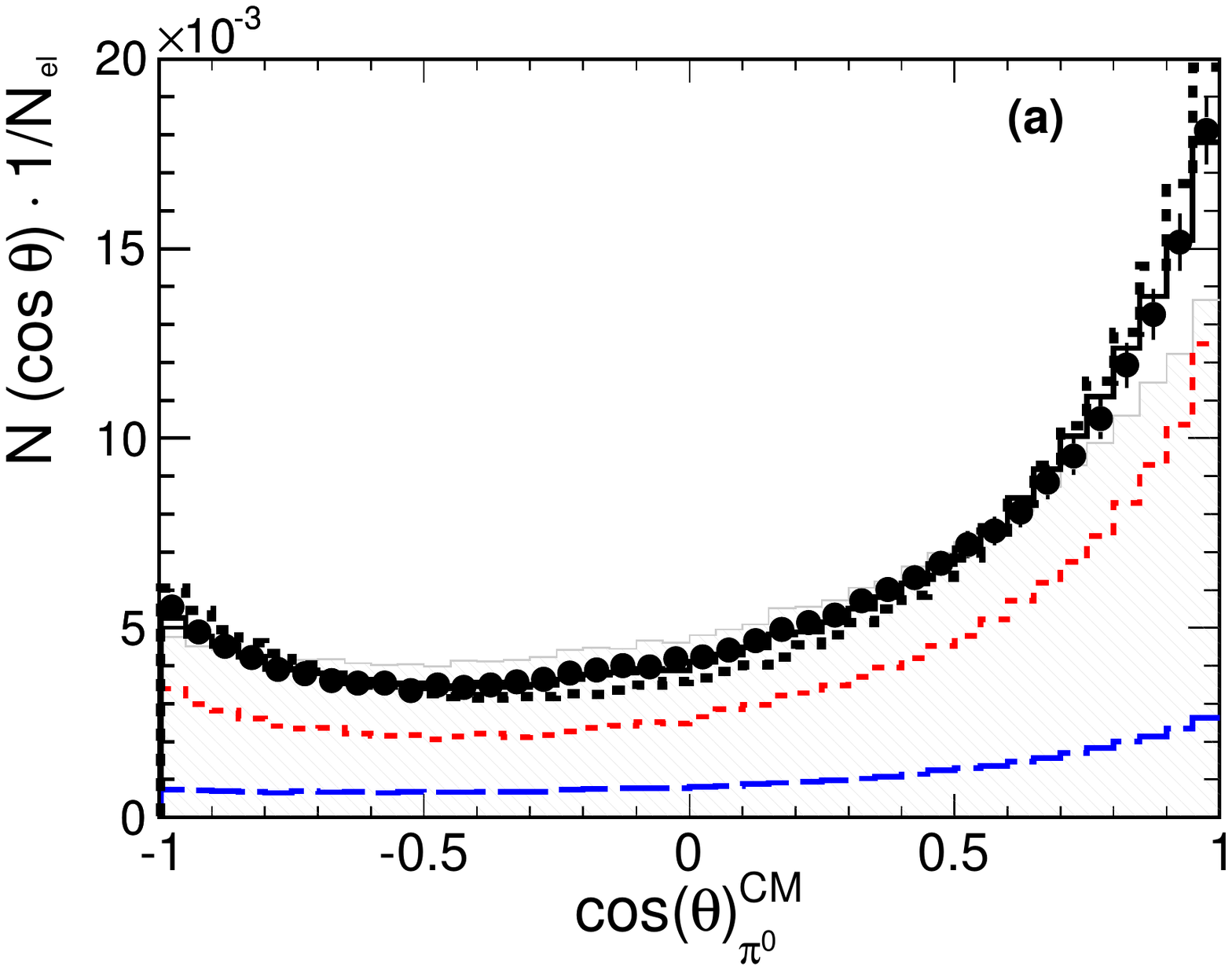}
                    \includegraphics[width=0.24\textwidth]{p_costh}
                    \includegraphics[width=0.24\textwidth]{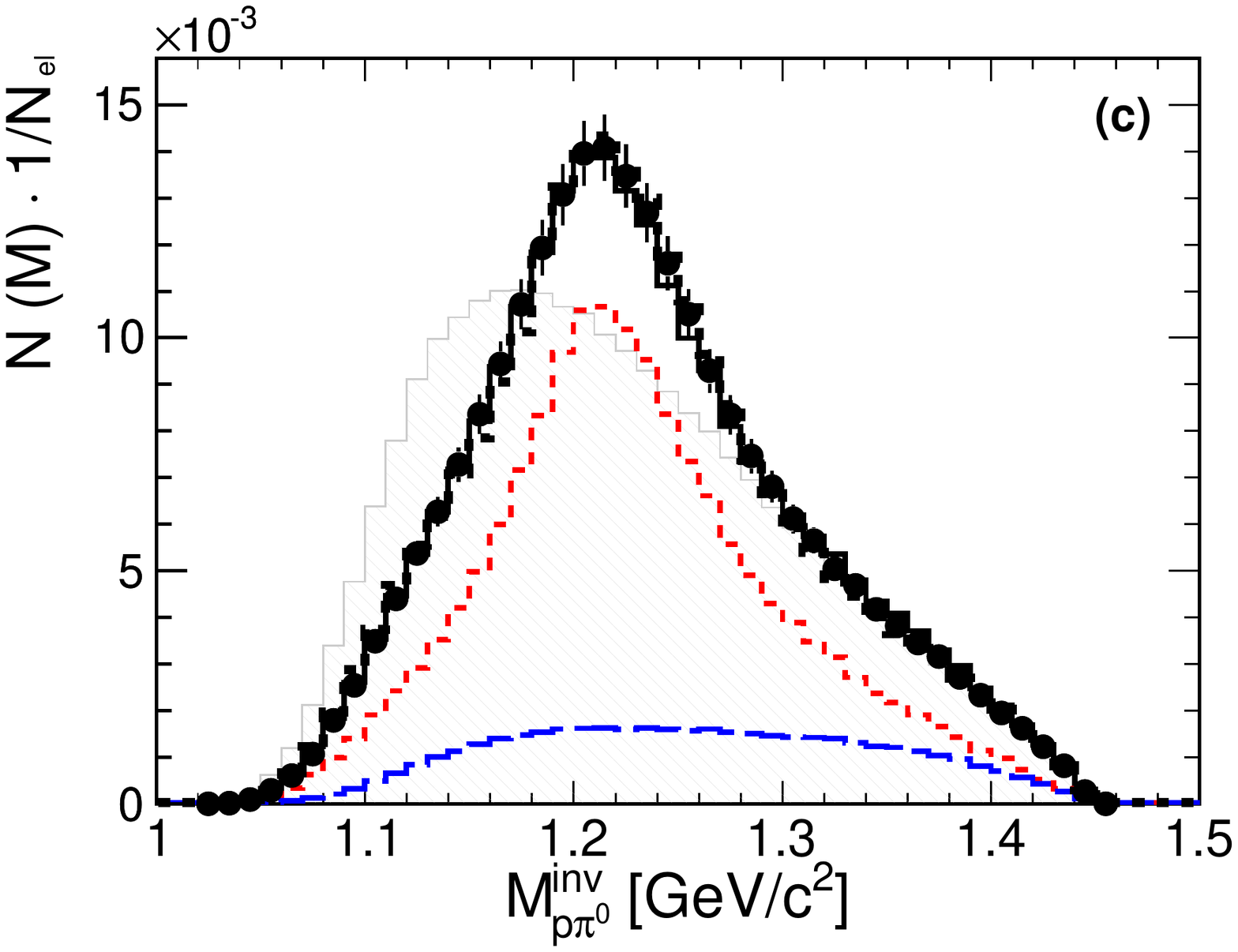}
                    \includegraphics[width=0.24\textwidth]{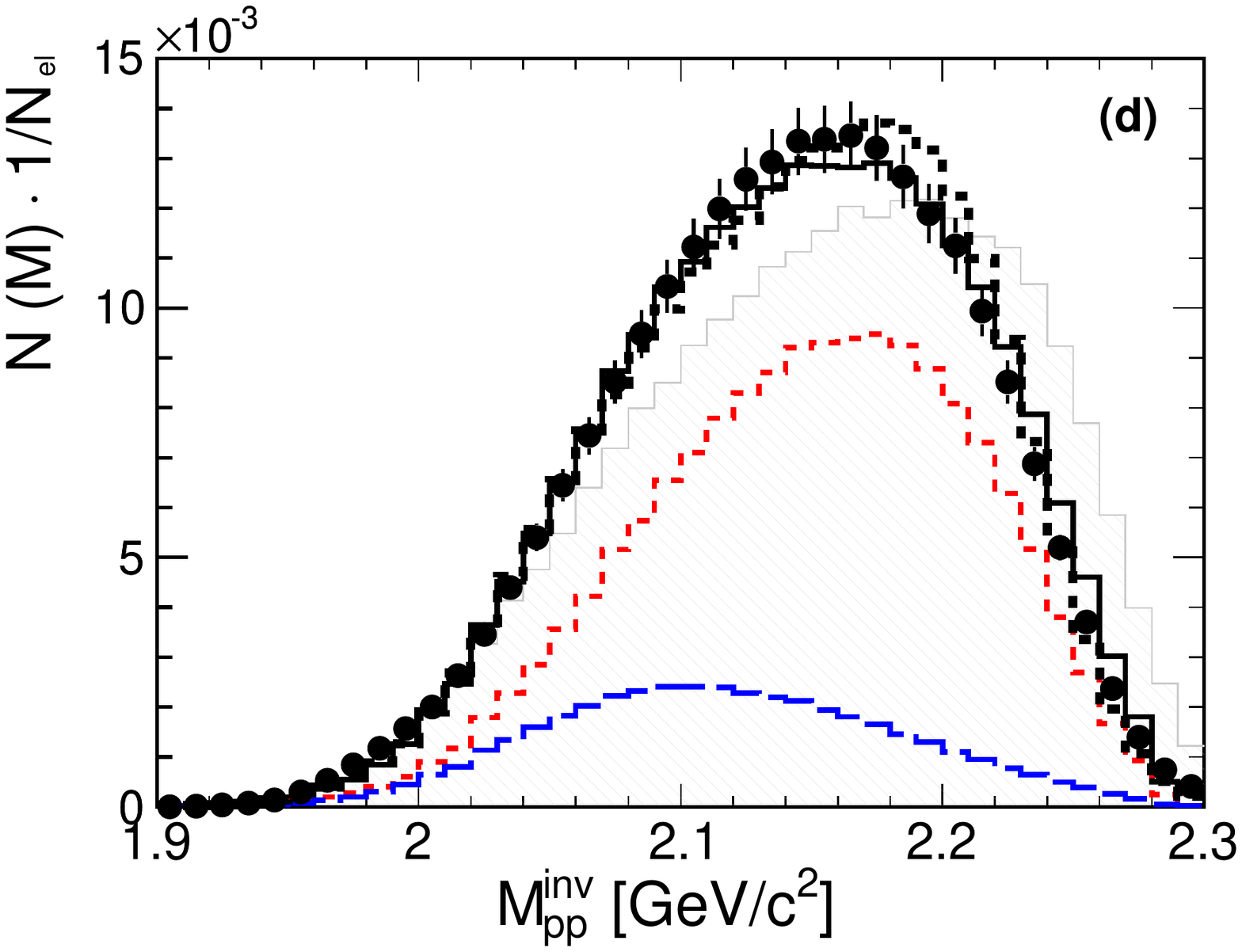}\\
                    Angular distribution of (a) $\pi^{0}$ and (b) $p$ in c.m.s. reference frame. Invariant mass of (c) $p\pi^{0}$ and (d) $pp$.\\

                    \includegraphics[width=0.24\textwidth]{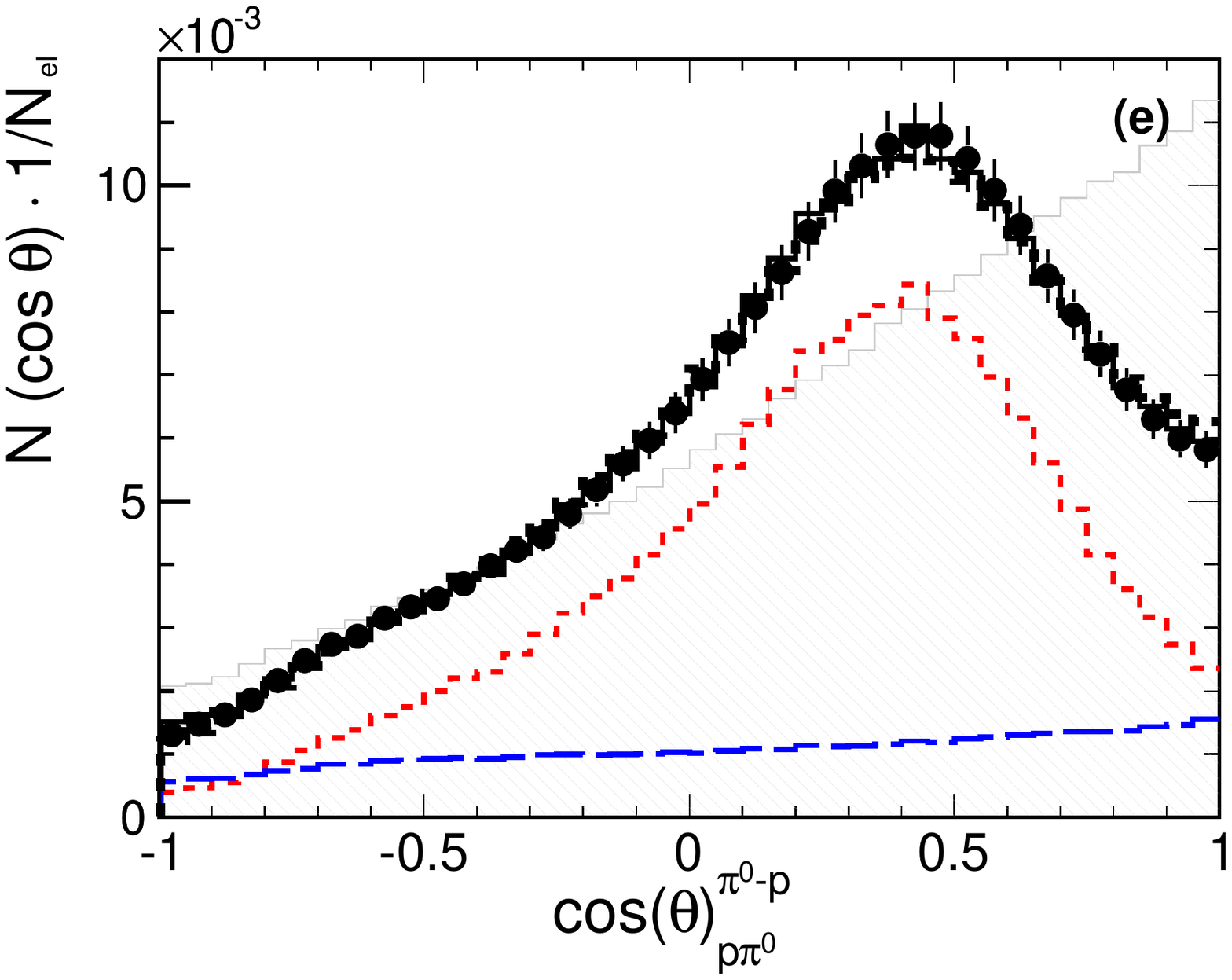}
                    \includegraphics[width=0.24\textwidth]{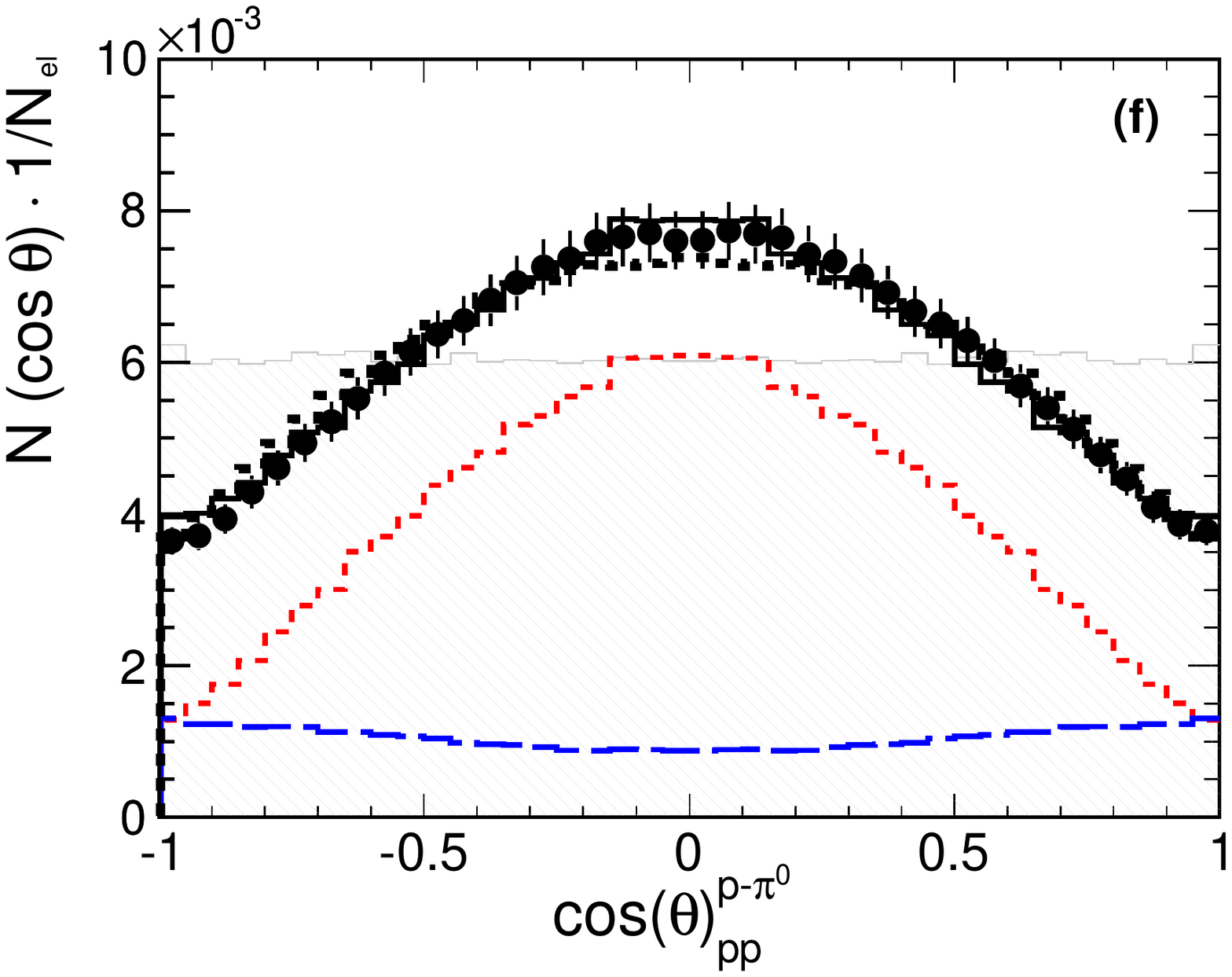}                    
                    \includegraphics[width=0.24\textwidth]{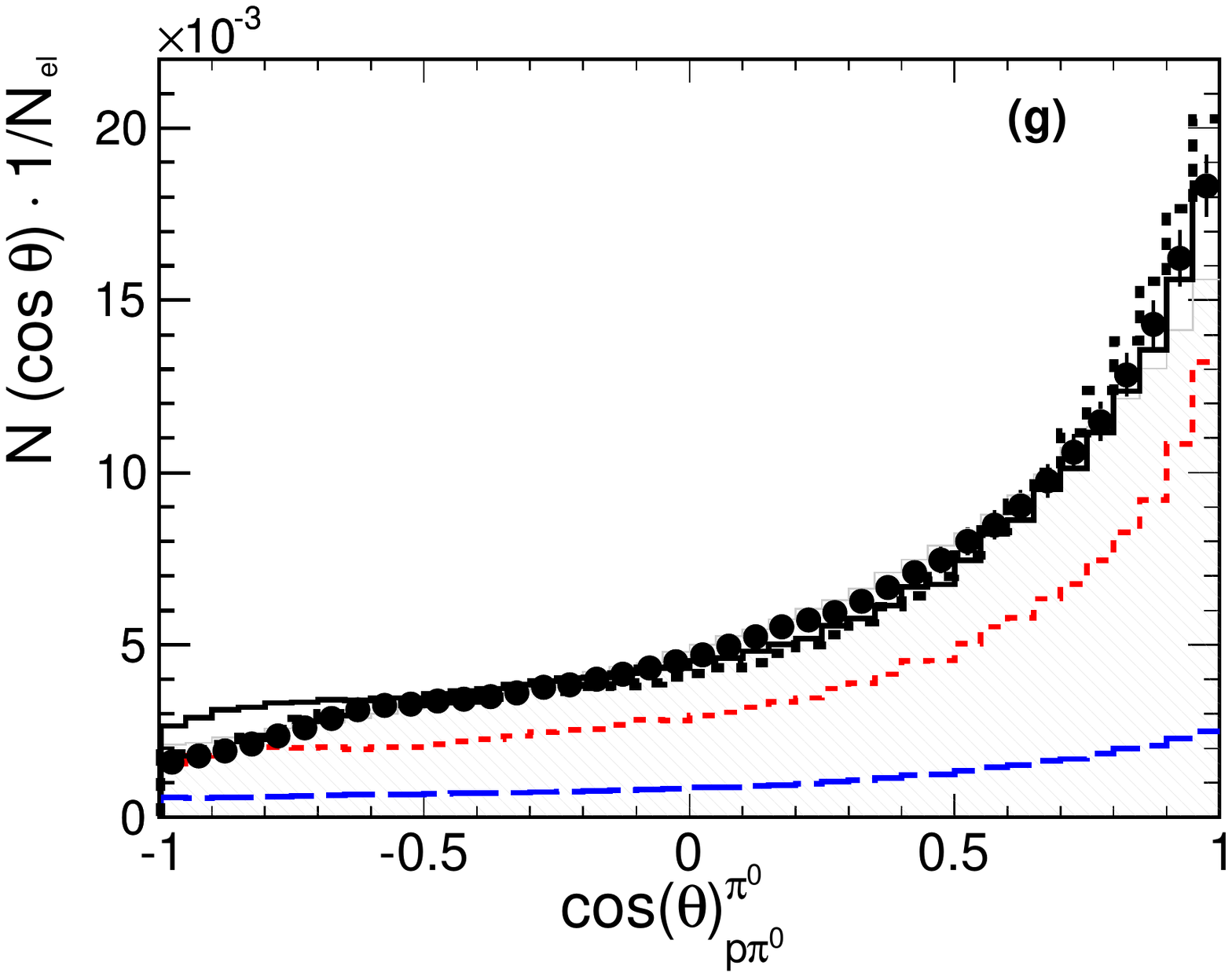}
                    \includegraphics[width=0.24\textwidth]{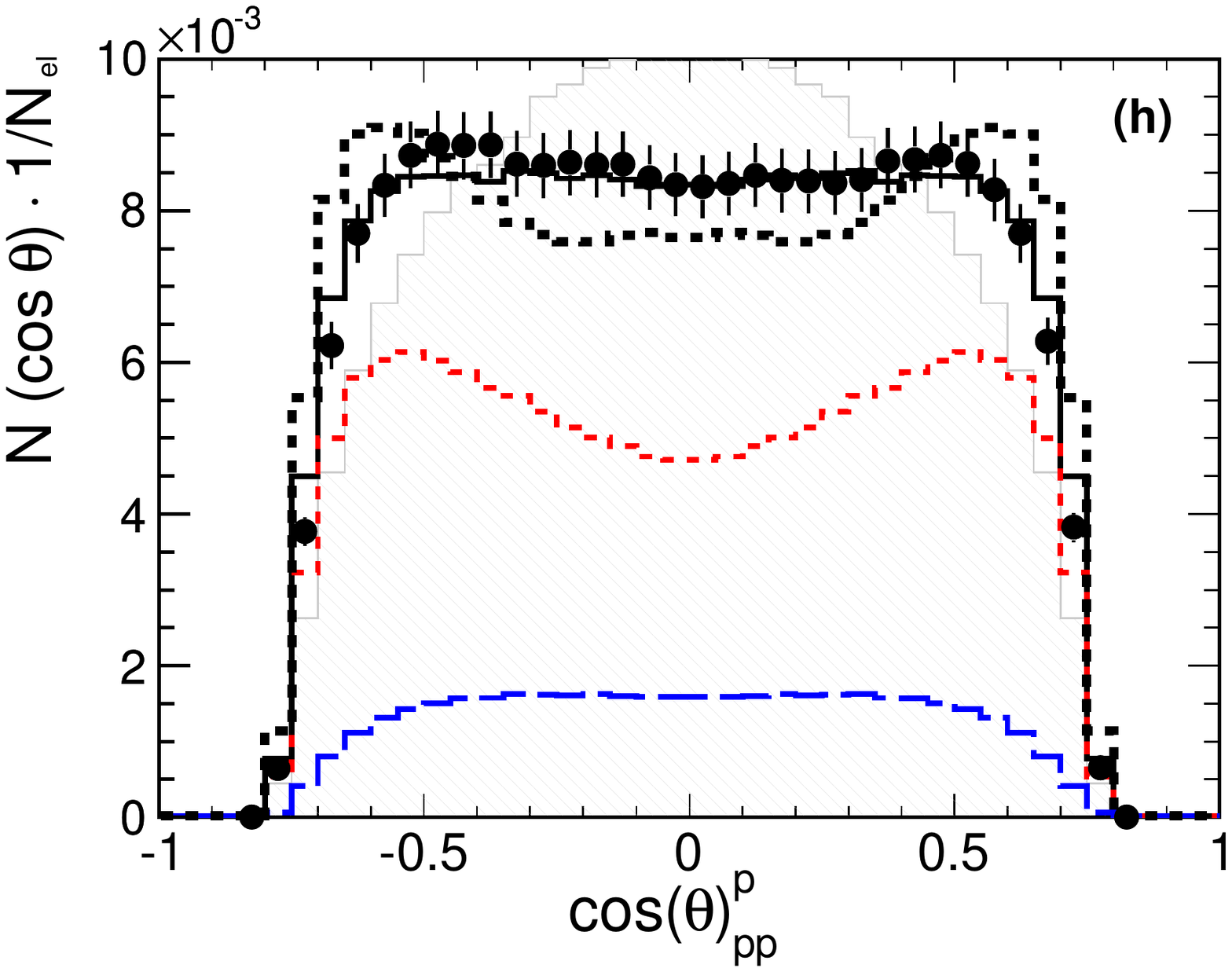}\\
                    Helicity distribution of (e) $\pi^{0}$ in $p\pi^{0}$ reference frame and (f) $p$ in $pp$ reference frame. Angular distribution of (g) $\pi^{0}$ in $p\pi^{0}$ GJ reference frame and (h) $p$ in $pp$ GJ reference frame.\\
            \caption{(Color online) Various projections for the $pp\pi^{0}$ channel:data points after acceptance corrections (black dots) based on the partial wave analysis solution. Data points in the areas of very low acceptance are omitted. Uncertainties originating from the various PWA solutions (as explained in the text) and statistical errors are visualized as grey band. Normalization error is not indicated. Histograms: total PWA solution (solid black), the $\Delta$(1232) contribution (short-dashed red) and the $N$(1440) contribution (long-dashed blue).}
    \label{f4}
    \end{figure*}

\begin{table}
\caption{The fitted data sets (number of events $N_{data}$).}
\begin{tabular}{lrrrr}
\hline
Reaction      &$\surd$s (MeV) &$N_{data}$ &$\sigma_{tot}$ (mb) & Reference \\
\hline
 $pp\to\pi^0pp$ & 2066    & 50000 &0.10\er0.03 &\cite{ElSamad2006} \\
 $pp\to\pi^0pp$ & 2157    &   542  &2.07\er0.09 & \cite{Andreev1994}   \\
 $pp\to\pi^0pp$ & 2178    &   615  &2.85\er0.13 & \cite{Andreev1994}   \\
 $pp\to\pi^0pp$ & 2200    &   882  &3.31\er0.19 & \cite{Andreev1994}   \\
 $pp\to\pi^0pp$ & 2217    &   993  &3.70\er0.14 & \cite{Andreev1994}   \\
 $pp\to\pi^0pp$ & 2234  &    914   &3.73\er0.15 & \cite{Andreev1994}   \\
 $pp\to\pi^0pp$ & 2251  &    996   &3.96\er0.15 &\cite{Andreev1994}   \\
 $pp\to\pi^0pp$ & 2269  &    1315  &4.20\er0.15 &\cite{Andreev1994}   \\
 $pp\to\pi^0pp$ & 2284  &    903   &4.19\er 0.17& \cite{Sarantsev2004}   \\
 $pp\to\pi^0pp$ & 2300  &    688   &4.48\er 0.20& \cite{Ermakov2011}   \\
 $pp\to\pi^0pp$ & 2319  &    1086  &4.50\er0.17 & \cite{Sarantsev2004}   \\
 $pp\to\pi^0pp$ & 2422  &   60000 &3.87\er0.55 & HADES \\
 \hline
 $pp\to\pi^+pn$ & 2285   &   4153  & 17.8\er0.4 &\cite{Ermakov2014} \\
 $pp\to\pi^+pn$ & 2300   &   2912  & 17.6\er0.6 &\cite{Ermakov2011}\\
 $pp\to\pi^+pn$ & 2422   &   60000 & 17.0\er2.2 &HADES \\
\hline
\end{tabular}
\label{data_base}
\end{table}

We started the analysis of the HADES data from the solution found in \cite{Ermakov2014}, describing low-energy data very well. The first fit produced a satisfactory description of the data, except of very forward neutron and very backward proton angles in c.m.s. of the $pp\to np\pi^+$ reaction. Moreover, we found large interferences between partial waves with Roper production and partial wave with non-resonant production of the $NN$ system. To stabilize the solution we included in the fit also the lower energy data fitted in \cite{Ermakov2014}. The fitted data base is given in Table~\ref{data_base}. Number of events $N_{data}$ used in the PWA is lower than the full available statistics in the case of \cite{ElSamad2006} (154972 events) and in the case of the HADES data (for the full statistics see Sec.~\ref{thehadesexp}).

To describe simultaneously the data in the energy range between $\sqrt{s}$=2.06 $GeV$ and $\sqrt{s}$=2.42 $GeV$, we introduce in the transition amplitudes a dependence on the total energy of the initial $NN$ system in the same form as in \cite{Ermakov2014,Ermakov2011}. Thus, the production of resonant and non-resonant two-body states was fitted by: 
\be
A_{tr}^\alpha(s)=\frac{a^\alpha_1+\sqrt{s}\,a^\alpha_3}{s-s^\alpha_0}\,e^{ia^\alpha_2},
\label{a1_a2_a3}
\ee
where $a^\alpha_1,a^\alpha_2,a^\alpha_3$ and $s^\alpha_0$ are real numbers, and the poles at $s=s^\alpha_0$ are located in the region of left-hand side singularities of the partial wave amplitudes. Indeed, in most of the fits, the only fitted function was the transition amplitude $A^\alpha_{tr}(s)$. In the case of transition from initial $NN$ state to a two-body state with stable particles, this function is a complex number at a fixed initial energy. In the case of the transition to the two-body subsystem (a resonance or non-resonant rescattering and a spectator) the transition amplitude has contributions from logarithmic singularities defined by the three-particle rescattering. Therefore, it should have a logarithmic dependence on the energy of the intermediate systems. However, this dependence  is not important for the production of such a relatively narrow state, as the $\Delta(1232)$ resonance. 
In the case of the Roper resonance we did not find a large difference between fits, where (i) the Roper total width was parameterized with the same energy dependence as the $\pi$N channel only or (ii) fits with a more complicated parameterization of the width with the following decay branching ratios: $\pi$N (60\%), $\Delta \pi$ (20\%) and N$(\pi\pi)_{S-wave}$ (20\%) (see \cite{Sarantsev2008}). We also made fits with free masses and widths of the $\Delta$ and Roper states. For $\Delta$(1232) the parameters hardly changed from the PDG values \cite{PDG2014}, and for the Roper resonance we only observed problems with convergence of the fit but no notable improvement of the description of the data. Extensive tests did not show any need for a more complicated energy dependence for the non-resonant amplitudes, either. All these solutions were included for our estimate of systematic errors.

Various solutions with the maximum total angular momentum $J=3$ or $J=4$ were considered. At first, we have performed the data base fit (see Table \ref{data_base}) with partial waves with total angular momentum up to $J=3$, since only these partial waves were found to be important for the description of the lower (than HADES) energy data \cite{Ermakov2014}. As in the case of the analysis of the HADES data alone such a fit describes rather well the $pp\to pp\pi^0$ single state but shows some problems in the description of the $pp\to np\pi^+$ reaction. In more details, the forward region of the neutron angular distribution calculated in c.m.s. of the reaction was underestimated by the fit. Let us mention that this angular region is completely covered by the HADES geometrical acceptance. As a consequence of such a description we obtained a rather small total cross section for the $pp\to np\pi^+$ reaction.

The very sharp behavior of the cross section at forward neutron angles (see Fig. \ref{f1}c) is reproduced well by the resonance model. This model includes an infinite number of partial waves based on one-pion exchange and, indeed, we should expect the largest contribution from high momentum waves at extreme angles. To check this idea we introduced in the Bonn-Gatchina analysis the partial waves with total angular momentum $J=4$ decaying into the $\Delta N$ intermediate state. A similar investigation was performed in \cite{Ermakov2014}. It was found that partial waves with the total angular momentum equal to four can contribute up to 6\% to the total cross section at highest energy (data set $\surd$s=2.3 GeV) but cannot be unambiguously identified. The present analysis produces a rather stable solution which defines the contribution from $J=4$ partial waves. It is found to be on the level of 5\% at $\surd$s=2.3 GeV in agreement with \cite{Ermakov2014} and it reaches 15\% at the HADES energies. Indeed, the fit with high spin partial waves reproduces rather well the forward angular distribution of the neutron in c.m.s. of the reaction (see Figs.~\ref{f1}c and \ref{f3}c). If partial waves with even higher $J=5$ are added to the fitting program we do not get an improvement of the solution but lose the convergence.

\begin{table}
\begin{center}
\begin{tabular}{|c|r|c|c|}
\hline
~& Total [\%] & $\Delta(1232)N$ [\%] & $N(1440)p$ [\%] \\
\hline
\multicolumn{4}{|c|}{$pp\to pp\pi^0$}\\
\hline
$^1S_0$ & 1.8  \er\, 0.7 & $<$1    & 1.8 \er\, 0.7\\
$^3P_0$ & 6.8  \er\, 1.0 & 1.5 \er\, 0.5 &5.5 \er\, 1.0 \\
$^3P_1$ & 21.0 \er\, 4.4 & 2.0 \er\, 1.0 & 12 \er\, 2.0  \\
$^3P_2$ & 29.5 \er\, 3.5 &30.5 \er\, 4.0 & 2.3 \er\, 1.0     \\
$^1D_2$ & 4.9  \er\, 1.0 & 4.2 \er\, 1.0 & $<$1         \\
$^3F_2$ &11.8  \er\, 2.0 & 6.5 \er\, 1.0 & $<$1         \\
$^3F_3$ & 2.0  \er\, 2.0 & 2.0 \er\, 2.0 & $<$1         \\
$^3F_4$ &12.0  \er\, 3.5 & 12.0 \er\, 3.0 & $<$1         \\
$^1G_4$ & 4.0  \er\, 1.0 & 4.0 \er\, 1.0 & $<$1         \\
$^3H_4$ & 5.5  \er\, 1.0 & 5.5 \er\, 1.0 & $<$1         \\
\hline
\multicolumn{4}{|c|}{$pp\to pn\pi^+$}\\
\hline
$^1S_0$ & 3.5  \er\, 0.8 & $<$1 & 2.2 \er\, 0.7\\
$^3P_0$ & 4.0  \er\, 1.5 & 1.0 \er\, 0.5 &1.7 \er\, 0.4 \\
$^3P_1$ & 14.0 \er\, 6.0 & 2.0 \er\, 1.0 & 6.7 \er\, 1.0  \\
$^3P_2$ & 33.5 \er\, 3.0 &29.5 \er\, 3.0 & 1.0 \er\, 0.5     \\
$^1D_2$ & 11.8  \er\, 1.5 & 8.8 \er\, 1.3 & $<$1         \\
$^3F_2$ &8.0  \er\, 1.0 & 6.5 \er\, 0.8 & $<$1         \\
$^3F_3$ & 2.0  \er\, 2.0 & 2.0 \er\, 2.0 & $<$1         \\
$^3F_4$ &11.5  \er\, 2.5 & 11.5 \er\, 2.5 & $<$1         \\
$^1G_4$ & 5.0  \er\, 1.0 & 5.0 \er\, 1.0 & $<$1         \\
$^3H_4$ & 5.5  \er\, 1.0 & 5.5 \er\, 1.0 & $<$1         \\
\hline
\end{tabular}
\caption{Contributions of the initial partial waves to the single pion production reaction $pp\to pp\pi^0$ and $pp\to np\pi^+$  at $\surd$s=2.42 GeV.}
\label{tablepwa}
\end{center}
\end{table}

The comparison of the measured data and Monte-Carlo events passed through the detector is shown in Figs. \ref{f1} and \ref{f2}. The PWA solution describes the data better than the one obtained with the modified resonance model and can be used for the acceptance correction of the HADES data. The acceptance corrected distributions are shown in Figs. \ref{f3} and \ref{f4}. The statistical errors are taken from the data, and model uncertainty errors are calculated from the set of solution described above (see the discussion below). Both statistical and model errors were added quadratically and are shown as a grey band.

\begin{figure*}
\centering
    \includegraphics[width=0.48\textwidth]{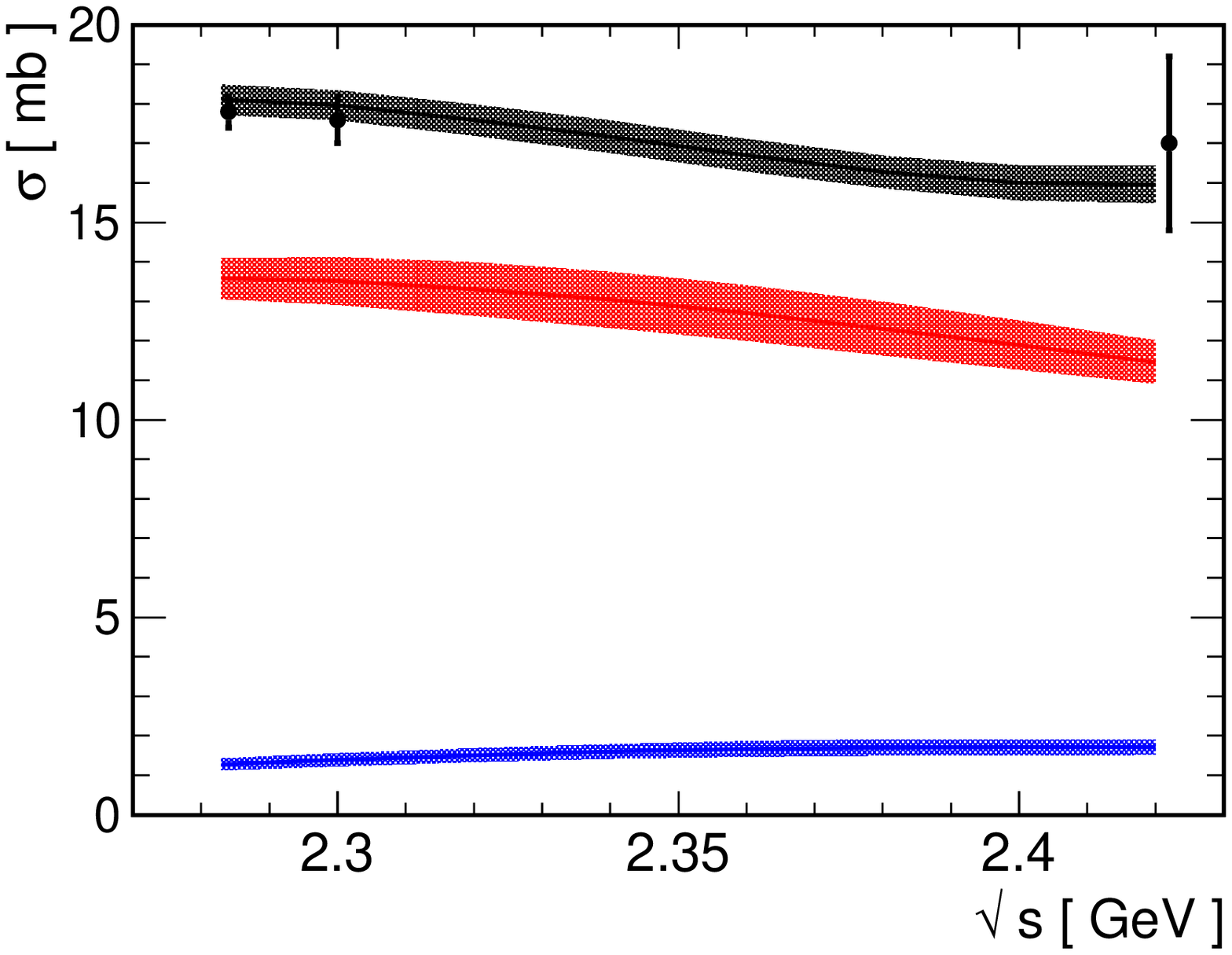}
    \includegraphics[width=0.48\textwidth]{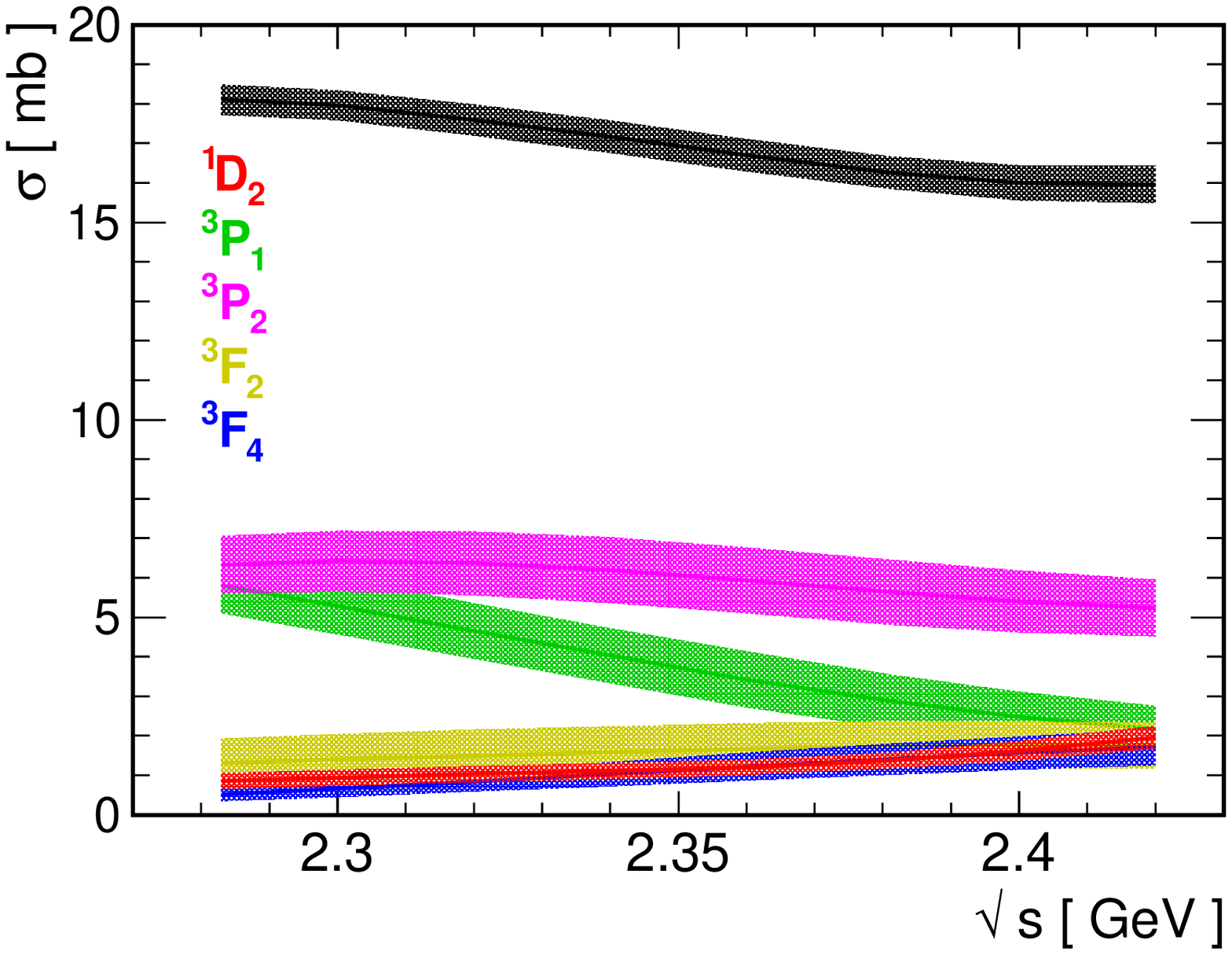}\\
  $np\pi^{+}$ channel\\
  
    \includegraphics[width=0.48\textwidth]{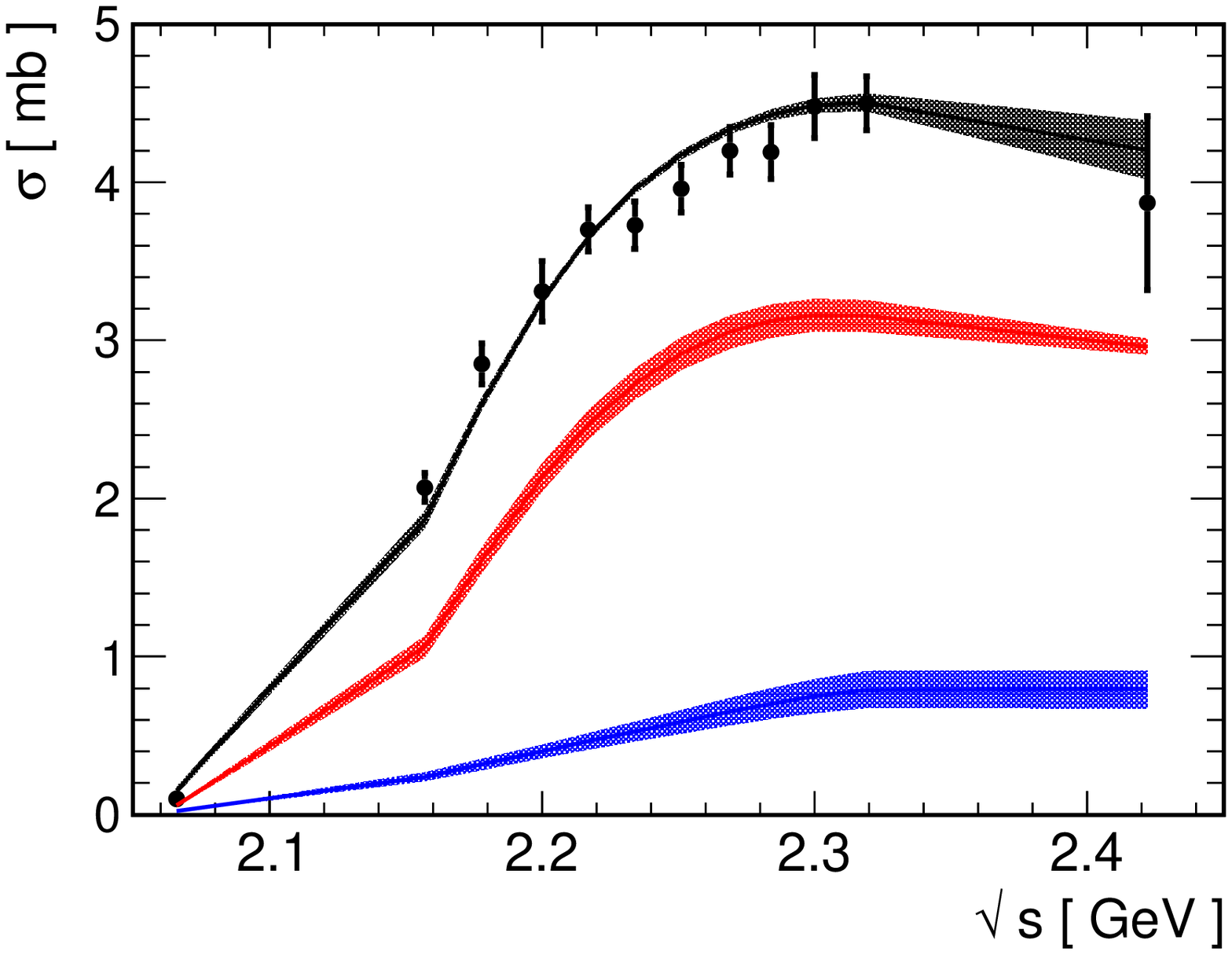}
    \includegraphics[width=0.48\textwidth]{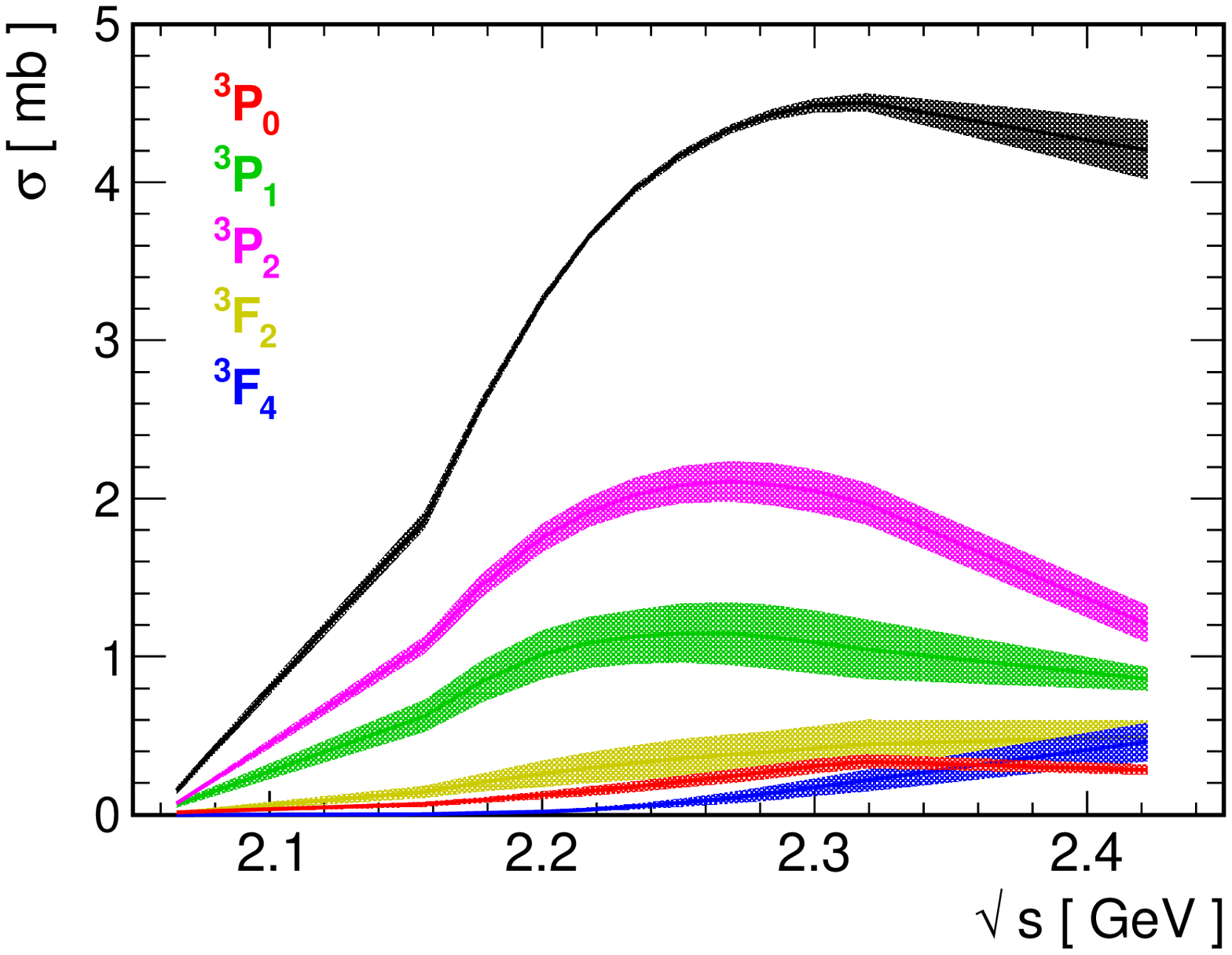}\\
  $pp\pi^{0}$ channel\\

\caption{(Color online) The description of the total cross section (data points with systematic error bars) in the combined analysis. Total partial wave solution (black curve) and contribution of $\Delta$(1232) (red) and $N$(1440) (blue) resonance in the PWA description (left), contributions of initial partial waves (right). Shaded bands reflect systematic uncertainties.}
\label{f5}      
\end{figure*}

Figure \ref{f5} shows the energy dependence of the pion production cross section ($np\pi^+$ upper panel, $pp\pi^0$ lower panel) and its decomposition into contributions of $\Delta$(1232), $N$(1440) (left) and incoming $pp$ partial waves (right). As expected, the cross section is dominated by the contributions from the partial waves with the $\Delta$(1232) resonance produced in the intermediate state. In the combined analysis of the data the partial waves with Roper production contributes about 20\% to the $pp\to pp\pi^0$ cross section and on the level of 12\% to the $pp\to np\pi^+$ cross section. The calculated contribution of the non-resonant terms amounts to 22-25\% in the $np\pi^+$ channel and 8-10\% in the $pp\pi^0$ channel.

Let us point out that the fit to the HADES data alone is optimized with a smaller Roper contributions: it was found to be around 10\% for $pp\to pp\pi^0$ and 6\% for the $pp\to pn\pi^+$ cross sections which is compatible with the modified resonance model results. Contrary to that model which includes an infinite number of the partial waves, the Bonn-Gatchina approach describes the data with a restricted number of partial waves (truncated method). It is based on an observation for the dominance of partial waves with low orbital momenta near production threshold. Thus, at the HADES energies, the amplitudes with production of a Roper state are included only with orbital momenta $L'=0,1$ between Roper and the spectator nucleon. Let us mention that the stability of the obtained solutions is tested by including in the fit partial waves with higher orbital momentum and checking that these contributions are small.

The contributions of the different initial partial waves to the HADES data as well as contributions of the partial waves with $\Delta$ and Roper production are listed in Table \ref{tablepwa}. The errors in Table~\ref{tablepwa} are defined from the set of solutions which include the combined fit of the whole data base, the fit of the HADES data alone and fits with contributions from higher spin states ($J=4$). In some of the fits we found notable interferences between non-resonant contributions in the $NN$ channel and Roper production. In the case of a large correlation we suppressed the non-resonant contributions and re-fitted the data. If the deterioration of the likelihood value was less than 1000 ($\sim$11$\%$) for the $pp\pi^0$ channel and less than 1500 ($\sim$4$\%$) for the $pn\pi^+$ channel and the fit did not show large systematic deviations in a particular distribution, it was also included in the error analysis. The uncertainties of both, initial partial waves and final state differential projections, span from the minimum to the maximum values obtained from the accepted set of the PWA solutions.

The same systematic approach was used for the calculation of errors for the total cross section obtained from the integration of the PWA solutions in the full solid angle. It was found to be 4.2~\er~0.15 mb for the $pp\to pp\pi^0$ reaction and 16.34~\er~0.8 mb for the $pp\to pn\pi^+$ reaction and the quoted errors are treated as the model uncertainty (see Table~\ref{cross_sections_tab} (column $\sigma_{PWA}$)). The correction of experimental data with the obtained PWA solution provides very similar cross section values: 4.1~\er~0.46 mb and 16.26~\er~1.96 mb, respectively. The errors, added quadratically, include: $5-6\%$ due to background subtraction and particle identification, $3-5\%$ the PWA model correction uncertainty and $8\%$ due to normalization. Both cross sections agree well within errors with the cross sections obtained with the modified resonance model approach. However the contribution of the partial waves with $\Delta$ production is smaller and there is a notable contribution from the non-resonant terms. These terms provide a rather stable common contribution but show a rather large variation between initial partial waves. The total cross section obtained in the partial wave analysis of all fitted data together with main contributions are shown in Fig.~\ref{f5} (right). The contributions from leading partial waves have a peak in the region slightly below 2.3 GeV. This peak is created due to a compromise between decreasing partial wave amplitudes and three-body phase volume which grows rapidly near the pion production threshold. A similar behavior was observed in the isospin $I=0$ sector \cite{Sarantsev2009}. It would be interesting to check whether such phenomenon can explain a resonance-like behavior of the $pn\to d\pi^+\pi^-$ total cross cross section observed in \cite{Kren2009}.

\section{Summary and conclusion}

The HADES data of the pion production reactions in proton-proton collision were analyzed with a modified OPE model and with the Bonn-Gatchina partial wave analysis method. A detailed study of various observables indicates that the partial wave solution provides not only a better control of the underlying physics but also a better description of experimental data (Figs. \ref{f1} and \ref{f2}). In the $pp\pi^{0}$ channel the discrepancies between PWA and the modified OPE model are visible in all spectra. Hence, the obtained PWA solution suits better to perform a full phase space acceptance correction of the measured data (Figs. \ref{f3} and \ref{f4}). 

The contribution of initial waves to the reactions cross section is defined as well as the contributions of partial waves with $\Delta$(1232) and Roper production in the intermediate state. The analysis shows that at given energy of $\surd$s=2.42 GeV the dominant contribution is defined by the production of $\Delta$(1232) in the intermediate state. This is visible not only in the proton-pion invariant mass distributions but also in the related helicity distributions. Furthermore, the pion angular distributions in the GJ frame shows a strong anisotropy, as expected from the $\Delta$ decay. The PWA solution attributes 75$\%$ of the total cross section to $\Delta$ in the $pp\to np\pi^+$ channel and 70$\%$ to $\Delta^{+}$ in the $pp\to pp\pi^0$ channel. Since no notable influence of the non-resonant partial waves was observed for the $\Delta$(1232) contribution, one can repartition the cross section for the $pp\to pp\pi^0$ reaction, obtaining the value $2.96\pm~0.22$ (syst.) $\pm~0.24$ (norm.) mb for the $\Delta$ resonance. The partial waves including the Roper production can contribute up to 20\% for $pp\to pp\pi^0$ and up to 12\% for $pp\to np\pi^+$.

\section*{Acknowledgments}

The HADES Collaboration gratefully acknowledges the support by the grants PTDC/FIS/113339/2009 LIP Coimbra, NCN 2013/10/M/ST2/00042 SIP JUC Cracow, 2013 /10/M/ST2/00042 Helmholtz Alliance HA216/EMMI GSI Darmstadt, VH-NG-823, Helmholtz Alliance HA216/EMMI TU Darmstadt, 283286, 05P12CRGHE HZDR Dresden, Helmholtz Alliance HA216/EMMI, HIC for FAIR (LOEWE), GSI F\&E Goethe-University, Frankfurt VH-NG-330, BMBF 06MT7180 TU M\"unchen, Garching BMBF:05P12RGGHM JLU Giessen, Giessen UCY/3411-23100, University Cyprus CNRS/IN2P3, IPN Orsay, Orsay MSMT LG 12007, AS CR M100481202, GACR 13-06759S NPI AS CR, Rez EU Contract No. HP3-283286. The work of A. V. Sarantsev and V. A. Nikonov is supported by RNF grant 14-22-00281.

\end{document}